\documentclass[prd,twocolumn,amsmath,nobibnotes,nofootinbib,superscriptaddress]{revtex4}

\usepackage{bm} \usepackage{graphicx}

\def\ba{\begin{eqnarray}}
\def\ea{\end{eqnarray}}
\def\be{\begin{equation}}
\def\ee{\end{equation}}
\def\({\left(}
\def\){\right)}
\def\[{\left[}
\def\]{\right]}
\def\<{\left<}
\def\>{\right>}

\begin{document}

\title{A status report on the observability of cosmic bubble collisions}
\date{\today}

\author{Anthony Aguirre}
\email{aguirre@scipp.ucsc.edu}
\affiliation{SCIPP, University of California, Santa Cruz, CA 95064, USA}
\author{Matthew C Johnson}
\email{mjohnson@theory.caltech.edu}
\affiliation{California Institute of Technology, Pasadena, CA 91125, USA}

\begin{abstract}
In the picture of eternal inflation as driven by a scalar potential with multiple minima, our observable universe resides inside one of many bubbles formed from transitions out of a false vacuum. These bubbles necessarily collide, upsetting the homogeneity and isotropy of our bubble interior, and possibly leading to detectable signatures in the observable portion of our bubble, potentially in the Cosmic Microwave Background or other precision cosmological probes. This constitutes a direct experimental test of eternal inflation and the landscape of string theory vacua. Assessing this possibility roughly splits into answering three questions:   What happens in a generic bubble collision? What observational effects might be expected? How likely are we to observe a collision? In this review we report the current progress on each of these questions, improve upon a few of the existing results, and attempt to lay out directions for future work.
\end{abstract}

\maketitle

\section{Introduction}
\label{sec-intro}

Cosmological inflation, the idea that the universe underwent a period of high-energy accelerated expansion, provides a natural and compelling explanation for the known hot, dense, nearly-homogeneous state of the universe 13.7 billion years ago. This explanatory power, combined with success in other precise cosmological tests (as well as a lack of similarly natural and compelling rival theories), has made inflation a key component of the `standard model' of cosmology accepted by many in the field.

In a great many specific realizations of this idea, however, inflation dramatically changes the picture of the universe on the largest scales because inflation is {\em everlasting}.  That is, inflation proceeds forever and ceases only {\em locally} in `pockets' or `bubbles' that may become radiation- or matter-dominated. One of these could contain our observable universe.

Everlasting (or `eternal') inflation can proceed in several ways (see, e.g.~\cite{Aguirre:2007gy,Guth:2007ng,Linde:2007fr} for recent reviews).  Here we focus on the picture of `false vacuum' (or perhaps `multi-minimum') eternal inflation.  This occurs in any inflaton scalar potential with multiple local minima, if the decay rate (via quantum tunneling) of a patch of space from one or more false vacua is sufficiently small that in one decay time the inflationary expansion more than doubles the patch's volume. In this picture, decay from the false vacuum can be mediated by the Coleman-DeLuccia (CDL) mechanism (see~\cite{Coleman:1980aw,Coleman:1977py}, and~\cite{Banks:2002nm} for a modern review), which if it can occur will be the most probable decay route.\footnote{This does not mean that CDL is necessarily how most transitions in an inflationary `landscape' proceed -- they may occur via the Hawking-Moss instanton (see~\cite{Hawking:1981fz}, and, e.g.,~\cite{Brown:2007sd,Batra:2006rz} for a modern view) or by some other means; but without a better picture of the form and statistical description of the potential landscape, the mechanisms' relative importance is difficult to gauge.}  Such a decay leads to the formation of an expanding bubble, the interior of which may be foliated by infinite, uniform, negatively-curved spatial sections -- a complete open universe!  Inflation may occur within this bubble, in which case the scenario has been dubbed `open inflation', and investigated extensively in the literature (see, e.g.,~\cite{Gott:1982zf,Linde:1995rv,GarciaBellido:1997uh,Garriga:1997ht,Linde:1999wv}).

Although this picture of a `multiverse' made up of bubbles is a natural side-effect of many models of inflation, it initially appears that bubbles other than ours are beyond any direct observational constraint, because each bubble is spatially infinite to observers inside.  Nonetheless, it has recently become apparent that {\em collisions} between bubbles just might leave relic observable signatures, and hence provide -- if we are very fortunate -- a direct way to test the eternal inflation scenario.

The study of bubble collisions has a fairly long history; indeed the fact that bubbles do not percolate (i.e. that sufficiently long-lived false vacua drive eternal inflation) was the fatal flaw in the first inflationary scenario (see, e.g.,~\cite{Guth:1982pn}).  The structure of bubble collisions was studied analytically and numerically in similarly early papers~\cite{Hawking:1981fz,Wu:1984eda}, and shortly thereafter Gott \& Statler~\cite{Gott:1984ps} proposed to constrain inflationary models by requiring that they {\em not} produce bubbles in our past lightcone, which were assumed to be incompatible with our observations.  The idea of open inflation then lay fallow for some time, with bubble collision studies primarily addressing gravity waves from early phase transitions (e.g.,~\cite{Kosowsky:1991ua,Turner:1992tz}).  After a brief resurgence during the mid-1990s (spurred by astronomical evidence for an open universe), false-vacuum eternal inflation returned to the fore with the emergence of `landscape' ideas in string theory.  In this context Refs.~\cite{Bousso:2006ge} and~\cite{Freivogel:2007fx} revisited the study of bubble collisions, finding exact solutions for vacuum bubbles (of non-positive vacuum energy) joined by a dust shell or domain wall.

The question of how many collisions might be expected to be in the past lightcone of an eternal-inflationary observer, left untouched since the early 80s, was taken up by Garriga, Guth \& Vilenkin~\cite{Garriga:2006hw} (hereafter GGV).  They found that this expected number depends upon the position inside the bubble, thereby defining a `center' to each bubble as well as a preferred frame for the overall bubble distribution.  This study too, however, implicitly assumed that {\em actual} observers cannot lie to the future of collision events.  A succeeding paper by the present authors and A. Shomer~\cite{Aguirre:2007an} (hereafter AJS)  supposed instead that observers {\em can} exist to the future of a collision, and asked: might they in fact actually observe signatures of the collision and thus test the eternal inflation picture?  That paper outlined the general conditions under which this would be possible, and calculated the number and angular size of collisions as they affect a very early-time surface within an observation bubble.

Following these papers have been a number of further works studying bubble collisions and their observability.  Nearly-simultaneous papers~\cite{Aguirre:2007wm,Chang:2007eq} generalized~\cite{Freivogel:2007fx} to calculate exact solutions for thin-wall de Sitter (dS) bubbles, with~\cite{Aguirre:2007wm} (hereafter AJ) also generalizing AJS to arbitrary cosmology for the observation bubble, and~\cite{Chang:2007eq} (hereafter CKL1) assessing some possible observable effects of the collision.  Following this, \cite{Chang:2008gj} (CKL2) deepened CKL1's discussion of observable effects; \cite{Aguirre:2008wy} (hereafter AJT) studied collisions using numerical simulations; Dahlen \cite{Dahlen:2008rd} studied collision probabilities from a slightly different perspective; \cite{Freivogel:2009it} (hereafter FKNS) revisited and generalized previous work on the statistics of bubble collisions. Finally, additional numerical work on bubble collisions was recently presented in Easther et. al.~\cite{Easther:2009ft}. 

While these papers represent enormous progress in understanding bubble collisions and their observability, the subject is quite subtle and complex. In this report, we have attempted to give a fairly comprehensive review of the subject that clarifies the general picture in terms of what is known, what is unknown, and what is as yet unclear.   
    
As a brief overview of the problem at hand, in order for us to hope to observe bubble collisions two basic conditions must hold. The first condition is that -- contrary to many years' assumptions -- bubble collisions must be such that observers like us might exist to the future of the collision event.  The studies listed above have revealed that collisions can have a large range of effects on a given set of physical observers. In order of decreasing severity, these might be classified and denoted as:
    \begin{enumerate}
    \item {\em Fatal}: collisions that either destroy, or prevent the formation of, physical observers to their future.\footnote{Here and in other cases, a collision might be fatal over some of its future and not the rest; this further complicates the probability assessment described below.}
    \item {\em Falsifiable:} collisions that potentially allow physical observers, but lead to a cosmology sufficiently different than ours that the corresponding scenarios can be easily ruled out.
    \item {\em Detectable:} collisions that allow observers who see a cosmology that is perturbed only slightly from the case with no collision. The signatures are detectable by present-day (or near-future) technology, and disentanglable from other effects.
    \item {\em Invisible:} As for `Detectable' collisions, but the signatures are too subtle to be measured by foreseeable technologies.
    \item {\em Non-existent:} collisions that are out of causal contact with the observer(s) in question.
    \end{enumerate}

Given that -- as found in previous work and reviewed in this report -- all five sorts of collisions are possible in principle, the focus falls on the second condition for collision observability: regions of spacetime in which observers exist and see collisions (Falsifiable or Detectable) must be reasonably {\em likely}, by some measure, relative to regions in which observers exist but don't see collisions (Non-existent or Invisible).

The problem of assessing these two conditions can be divided into three intertwined sub-problems, progress in which we shall review in this paper. First, we can ask what spacetime and field structure results from a generic collision between two bubbles and how it depends on the two bubbles' properties, nucleation locations, field potential, etc.; this question is addressed in Sec.~\ref{sec-genericcollision}, after a review in Sec.~\ref{sec-onebubreview} of the structure of, and cosmology inside, a single bubble.  Second, we can assess what potential observables exist for a given observer to the future of a bubble collisions; this is treated in Sec.~\ref{sec:obssignatures}. Third, placing such collisions in a global picture of eternal inflation, we can ask for the probability distribution of different levels of effect (as categorized above) for a `randomly chosen'~\footnote{Aficionados of the multiverse, among others, will realize that this phrase is fraught with subtleties, to say the least.} physical observer; Section~\ref{sec-probabilities} reviews progress on this question. Note that although this is primarily a review, a number of new results are included, which are identified as we proceed. Notation is generally chosen to be compatible with that of AJS, AJ and AJT; except where noted, we work in natural units. 

\section{The universe in a nut shell}
\label{sec-onebubreview}

In this section, we review the structure of bubble universes formed during false vacuum eternal inflation in 3+1 dimensions. Consider a theory with a set of scalar fields and associated potential $V$ with positive energy local minima. Transitions out of these unstable false vacua can be mediated by instanton solutions with probability per unit four volume~\cite{Coleman:1977py}
\begin{equation}\label{eq:decayrate}
\lambda = A \exp(-S_{E}),
\end{equation}
where $A$ is a pre-factor of mass dimension four encoding the first quantum corrections~\cite{Callan:1977pt}, and $S_{E}$ is the Euclidean action of the instanton solution. If it exists, the Coleman de-Luccia (CDL) instanton~\cite{Coleman:1980aw} (which is the maximally symmetric compact solution, possessing $O(4)$ invariance) will have the lowest action. This instanton has metric
\begin{equation}\label{eq:instantonmetric}
ds^2 = d\tau^2 + a(\tau)^2 d\Omega_3^2,
\end{equation}
where $\tau$ has a finite range, over which the three-sphere's scale factor $a$ evolves between two zeros. The scale factor is coupled via the Euclidean Einstein equations to the fields, which are in turn governed by the inverted potential $-V$. During the metric's evolution between zeros of the scale factor, the fields interpolate between two turning points, one in the basin of attraction of the false vacuum, and the other outside of it. We assume that all transitions are mediated by the CDL instanton in the following. 

\subsection{Nucleation rates}
The Euclidean action is determined by the properties of the scalar potential, and there are a number of important cases to distinguish. Including the effects of gravity, the action is composed of two pieces:
\begin{equation}
S_{E} = S_{\rm inst} - S_{BG},
\end{equation}
where $S_{\rm inst}$ is the action of instanton solution itself, and $S_{BG} = - 24 \pi^2/V_F$ is a background subtraction determined by the energy density $V_F$ of the false vacuum. The background subtraction is large unless $V_F \sim 1$. We will assume (so that these semi-classical arguments make sense) that the false vacuum energy scale is somewhat lower than the Planck scale, in which case $S_E \gg 1$  unless $S_{\rm inst}$ and $S_{BG}$ cancel very precisely. This can happen when the potential satisfies what is known as the thin-wall approximation, where the potential difference $\Delta V$ between the true and false vacua satisfies $\Delta V \ll V_F$. Within this approximation, the field `loiters' near the false vacuum for most of the range in $\tau$, during which time the instanton Eq.~\ref{eq:instantonmetric} resembles the Euclideanized false vacuum. The field then quickly traverses field-space and loiters for some shorter interval in $\tau$ near the true vacuum. The instanton solution can then be thought of as two four--spheres matched across a thin wall of some tension $\sigma$. 

The radius at which the wall occurs (which will be equal to the nucleation radius of the Lorentzian bubble solution discussed below) is determined by the tension and difference in vacuum energy at the instanton endpoints. When the initial radius $R_{\rm nuc}$ is much smaller than the false vacuum horizon size $H_F^{-1}$, gravitational effects are not important in determining the radius and action, which are then given by
\begin{equation}
R_{\rm nuc}= \frac{\sigma}{\Delta V}, \ \ \ \ S_E = \frac{27 \pi^2 \sigma^4 }{2 (\Delta V)^3}.
\end{equation}

Unless the energy scales in the potential are set by $m_p$ (so that even outside the thin wall approximation the action is small) or gravitational effects are unimportant (so that there can be a precise cancellation between the instanton action and background subtraction), $S_E$ will typically be much larger than one. Since the natural scale determining the pre-factor is $H_F$, this suggests that we expect $\lambda H_F^{-4} \ll 1$. For such small nucleation rates, less than one bubble nucleation event will occur per false vacuum Hubble four-volume, and false vacuum eternal inflation will ensue. Estimating exactly {\em how} suppressed one should expect the rates to be is, however, quite difficult given our lack of knowledge about the origin of the underlying inflaton potential. 

\subsection{Cosmology inside the bubble}
\label{sec-cosmobubble}

\begin{figure*}[htb]
\includegraphics[width=14 cm]{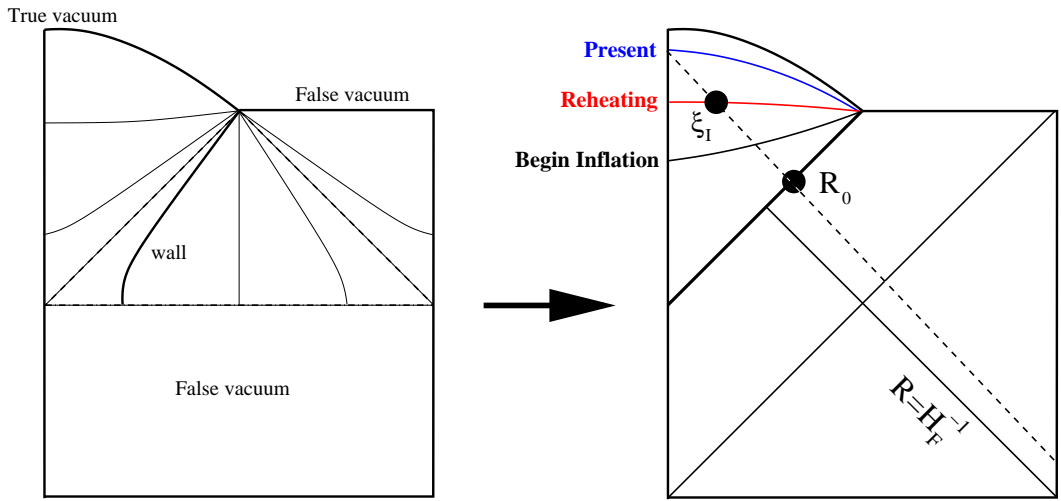}
\caption{On the left, we depict the post-nucleation spacetime. Surfaces of constant field in each of the three coordinate regions (the true vacuum on the left, wall region in the center, and false vacuum on the right) are drawn as solid lines. The coordinate regions are separated by the null dashed lines, on which the field is at the instanton endpoints. If the instanton is thin-wall, the field jumps between the instanton endpoints at a fixed value of $\tau$, which is identified as the wall (dark solid line). If in addition the initial radius of the bubble is small, it is possible to approximate the bubble wall as lying on the light cone as shown in the right panel. Outside of the bubble is nearly pure false vacuum, and inside is an open FLRW universe. Surfaces of constant density are depicted by the solid colored lines, which correspond to various cosmological epochs. Following the past light cone of a present-day observer back, it will intersect the surface of last scattering at $\xi = \xi_{\rm ls}$, and the bubble wall when it has radius $R_0$ (given by Eq.~\ref{Rosimp}).
 \label{fig-bubbleanatomy}
}
\end{figure*}

The CDL instanton can be analytically continued in order to obtain the post-nucleation Lorentzian spacetime containing a bubble, depicted in the left panel of Fig.~\ref{fig-bubbleanatomy}. The analytic continuation of the metric Eq.~\ref{eq:instantonmetric} yields three regions separated by coordinate singularities. Continuing one of the angles of the three-sphere, $\theta \rightarrow i \xi + 3 \pi / 2$, gives the spacetime and field configuration in the vicinity of the wall. In this region, the field interpolates between the turning points of the field evolution (described above), i.e. between a point near the false vacuum and the point at which the field `emerges' after tunneling through the barrier.  In cases where the thin-wall approximation holds, the field jumps between its two endpoints at a specific value of $\tau$; this jumping region can be described as a timelike wall of tension $\sigma$, which because of the hyperbolic nature of the slices of constant $\tau$, can be viewed as undergoing constant acceleration. In the remainder of the paper, we will generally assume that 
\begin{itemize}
\item The thin-wall approximation holds.
\item The small-bubble approximation, $H_F R_{\rm nuc}\ll 1$, holds.
\end{itemize}

The region just described is bounded by a coordinate singularity at the Euclidean time endpoints $\tau=0$ and $\tau=\tau_{\rm max}$, at which $a = 0$ and all field derivatives are zero. The configurations on the other side of these coordinate singularities can be obtained from Eq.~\ref{eq:instantonmetric} by taking $\tau \rightarrow i \tau$ and $\theta \rightarrow i \xi$, each of which yields a region in which the metric is that of an open Freidmann-Lema\^itre-Robertson-Walker (FLRW) universe:
\begin{equation}\label{opends}
ds^2 = -d \tau^2 +a^2(\tau) \left[d \xi^2 + \sinh^2 \xi \ d\Omega_{2}^{2} \right].
\end{equation}
One can also define a conformal time $\Xi$ by
\begin{equation}
d\Xi=d\tau/a(t),
\end{equation}
giving metric:
\begin{equation}
ds^2 = a^2(\tau) \left[- d \Xi^2 + d \xi^2 + \sinh^2 \xi \ d\Omega_{2}^{2} \right].
\end{equation}
Since all field velocities vanish on the $a=0$ `big-bang' surfaces, there are no curvature singularities, and each FLRW universe begins with an epoch of curvature domination. In one FLRW region, the field relaxes back to the false vacuum, and in the other it rolls down to the true vacuum. 

Depending on the content of the theory, there can be interesting cosmological evolution as the field evolves away from the endpoints.  We will assume the following cosmology inside of the bubble:
\begin{enumerate}
\item At $\tau = 0$, all field derivatives are zero, and the universe is curvature dominated. The scale factor is given by
\begin{equation}\label{eq:ainflation}
a(\tau) = H_I^{-1} \sinh (H_I \tau)
\end{equation}
where 
\begin{equation}
H_I = \sqrt{\frac{V_I}{3}}.
\end{equation}
 (This is generalized in Appendix~\ref{sec:bubblecosmoapp} to the case where inflation starts after an arbitrarily long initial epoch of curvature domination, as could occur if the field tunneled into a phase of fast-roll.)
\item Inflation occurs at scale $H_I$ for $N_e = H_I \tau_{\rm reh}$ e-foldings. 
\item The universe instantaneously reheats at $\tau = \tau_{\rm reh}$ into an epoch of radiation domination, followed by matter domination. The scale factor can be written in terms of the conformal time as (see e.g.~\cite{Mukhanov:2005sc}): 
\begin{equation}\label{eq:aradmat}
a (\Xi) = a_* \left[ \sinh \Xi + r \left( \cosh \Xi - 1 \right) \right]
\end{equation}
where
\begin{equation}
r = \frac{1 - \Omega_c - \Omega_r }{2 \Omega_r^{1/2} \Omega_c^{1/2}},\ \ 
a_* = \frac{\Omega_r^{1/2}}{H_0 \Omega_c}, 
\end{equation}
with $\Omega_r$ the current fraction of the critical density in radiation, $\Omega_c$ that in curvature, and $H_0$ the present Hubble constant. Matter and radiation equality occurs at $a_{\rm eq} = a_* / 2 r$. 

\item At late times, the energy density becomes dominated by a positive cosmological constant.
\end{enumerate}

In the thin-wall and small-bubble approximations we make, the one-bubble spacetime is shown in the right panel of Fig.~\ref{fig-bubbleanatomy}. There are a number of important quantities necessary to connect a realistic inner-bubble cosmology with the effects of bubble collisions, which we list below and elaborate upon in Appendix~\ref{sec:bubblecosmoapp}: 
\begin{enumerate}
\item We require the minimal number of inflationary efolds ($N_e$) to be consistent with current bounds on curvature $\Omega_c$. Consistency with the WMAP 5-year data~\cite{Komatsu:2008hk} ($\Omega_c < .0081$ within $95 \%$ confidence level and $\Omega_r \simeq 8 \times 10^{-5}$) requires
\begin{equation}
N_e > {\rm arccosh} \left[{1\over 2} \frac{T_{\rm reh}}{10^{-11} {\rm GeV}} \right],
\end{equation}
For reheating temperatures $T_{\rm reh}$ ranging from $1\, {\rm TeV} \leq T_{\rm reh} \leq 10^{16}\, {\rm GeV}$ corresponds to $30 \leq N_e \leq 62$. 
\item We require the distance in $\xi$ on earlier $\tau$ surfaces out to which an observer has causal access, which we denote as $\xi_{\rm view}$. Of most relevance is the distance $\xi_{\rm reh}$ on the reheating surface and the distance $\xi_{\rm ls}$ on the surface of last scattering. Because null rays obey $d\Xi = \pm d\xi$, and because the conformal time is quite small at both the time of reheating and last scattering, $\xi_{\rm ls, reh}$ is approximately given by the observer's conformal time. If we are the observer, this can be obtained setting Eq.~\ref{eq:aradmat} to one and solving for $\Xi$. Using the previously mentioned bounds from the WMAP 5-year data,
\begin{equation}\label{eq:xireh}
\xi_{\rm ls} \simeq 2 \Omega_c^{1/2} < 0.18.
\end{equation}
\item We require the radius $R_0$ of the bubble wall (which in the small bubble approximation is the $\tau = 0$ surface) at its intersection with the past light cone of a present-day observer at time $\tau_o$. Neglecting the late-time epoch of vacuum energy domination (which only adds a small correction), we find:
\begin{equation} \label{Rosimp}
R_0 = H_I^{-1} \tanh \left( \frac{N}{2} \right) \left[ \frac{1+ \Omega_{r}^{1/2} + \Omega_c^{1/2}}{1+ \Omega_{r}^{1/2} - \Omega_c^{1/2}} \right].
\end{equation}
An enhancement is only possible if there were to be a longer period of curvature domination immediately following the formation of the bubble.  As shown in Appendix~\ref{sec:bubblecosmoapp}, if a period of curvature domination of duration $\tau_c \agt H_I^{-1}$ precedes inflation in the bubble, $R_0$ scales up by a factor of approximately $\tau_c H_I^{-1}$.
\end{enumerate}
These quantities, which describe different aspects of matching the bubble interior to the false vacuum exterior, are labeled in Fig.~\ref{fig-bubbleanatomy}. 

\section{What happens in a generic bubble collision?}
\label{sec-genericcollision}
The single-bubble model of the previous section provides a viable cosmology. However, individual bubbles do not exist in complete isolation: as they grow, each bubble undergoes an unlimited number of collisions with other bubbles nucleated from the same false vacuum background. This section analyzes what happens in a collision between two vacuum bubbles. Three or more colliding bubbles is interesting, but two is sufficient for assessing the generic outcome of a collisions and also allows a number of simplifications, as follows.

We assume, without loss of generality, that one of the bubbles is centered around the origin of coordinates in the background false vacuum dS. A single bubble's SO(3,1) symmetry then means that `boosts' about the origin do not change the bubble configuration. Adding another bubble breaks this symmetry, but in this case a boost can be used to go to a frame in which both bubbles are nucleated at the same time in the background dS. (We will comment further about such boosts in Sec.~\ref{sec-refframes}, where we refer to this frame as the `collision frame,' and defer further details until then.) In the collision spacetime, a residual SO(2,1) hyperbolic symmetry is retained, owing to the fact that the intersection of the two SO(3,1) symmetric field configurations is always a hyperboloid itself. This symmetry allows for a drastic simplification in description of the collision dynamics by rendering the problem $1+1$ dimensional. 

The sole kinematical variable in this setup is the initial separation of the bubbles; all of the other properties of the collision are determined by the underlying scalar potential, about which some set of assumptions and approximations must be invoked to make progress. The simplest model has a single true and false vacuum, so that only identical bubbles undergo collisions. In this case, because the bubbles have interiors of the same phase, they simply merge. When more than one bubble-type can be nucleated out of the same false vacuum, collisions occur between all bubble types. When bubbles containing different vacua collide, they cannot merge, and a domain wall must form after the collision.

In addition to a possible post-collision domain wall, there generally must be debris released by the collision in order to conserve energy and momentum. The exact nature of this debris is dependent upon the couplings and field content of the underlying theory. It was observed in early numerical simulations~\cite{Hawking:1982ga}, and later in~\cite{BlancoPillado:2003hq,Aguirre:2008wy,Easther:2009ft}, that the debris in at least some cases are released as a null `shell' from the location of the collision. 

A reasonable model of the collision therefore includes a possible post-collision domain wall, and debris emitted in a null shell. Discounting any internal degrees of freedom of the debris shell and domain wall, and further assuming that the colliding bubbles each satisfy the thin-wall approximation, it is possible to describe the collision as an Israel junction condition problem between a number of spacetime regions. This approach was first adopted by Wu in Ref.~\cite{Wu:1984eda} for the collision between two identical bubbles of zero vacuum energy, and has been used in nearly all subsequent analytic studies of bubble collisions~\cite{Bucher:2001it,BlancoPillado:2003hq,Bousso:2006ge,Chang:2007eq,Freivogel:2007fx,Aguirre:2007wm}. We will summarize some of the important elements of this picture in Sec.~\ref{sec:thinwallcollisions}.

While useful for determining the global structure of a collision spacetime, thin-wall matching misses a number of properties that will be key for assessing the observability of collisions. For example, because the matching is done between vacuum solutions, it is impossible to obtain information about the effect of collisions on the single-bubble open FLRW foliation. This knowledge is critical since the foliation of the post-collision spacetime into surfaces of constant density will determine not only the relevant set of observers, but also what they will see. This question was addressed using semi-analytic methods in CKL2 and using numerical simulations in AJT; we describe some of the important conclusions of these studies in Sec.~\ref{sec:numericalsims}.

\subsection{Results from exact solutions of thin walls between vacuum bubbles}\label{sec:thinwallcollisions}

As mentioned above, the basic features of a collision spacetime can be captured by matching spacetime regions across a post-collision domain wall and null shells of collision debris. Earlier papers treated the collision of identical zero vacuum energy bubbles~\cite{Wu:1984eda,Bousso:2006ge}, zero and negative vacuum energy bubbles~\cite{Freivogel:2007fx}, and the general case of negative, zero, or positive vacuum energy bubbles~\cite{Aguirre:2007wm,Chang:2007eq}. In this section, we outline the assumptions made in the thin-wall matching problem, and highlight a number of important results obtained from this analysis. We refer the reader to AJ and CKL1 for further details on this construction.

A treatment applicable to observable bubble collisions involves the following set of assumptions:
\begin{itemize}
\item The background space time is dS, with Hubble constant $H_F$.
\item The `observation' bubble is (prior to the collision) either dS or Minkowski, with Hubble constant $H_o$, and nucleates with proper radius $R_{\rm nuc,o}$.  It is joined to the background by a thin domain wall of tension $\sigma_o$. 
\item The `collision' bubble is dS, AdS or Minkowski, with Hubble constant\footnote{Note that using our conventions, $H^2=0$ for Minkowski and $H^2 < 0$ for AdS.} $H_C$, nucleated with proper radius $R_{\rm nuc,C}$, and joined to the background by a wall of tension $\sigma_C$. 
\item When the two vacua inside of the colliding bubbles are distinct, a domain wall of tension $\sigma_{oC}$ will form after the collision. To conserve energy-momentum an extra energy sink is required, which is modeled as two outgoing null shells of some initial (possibly different) energy density.~\footnote{For some types of colliding bubbles and kinematics, there may be additional dynamics in the vicinity of the collision. One example is the collapsing and expanding pockets of false vacuum observed in the early simulations of~\cite{Hawking:1982ga}. Another example is the classical transition mechanism of Easther et. al., which we describe in the next section. Future work could extend the thin-wall analysis to these situations.}
\item The collision spacetime can be constructed by matching vacuum solutions with SO(2,1) symmetry across the post-collision domain wall and radiation shells.
\end{itemize}

There are a total of five different spacetime regions to match: the false vacuum and two true vacua, then the post-collisions spacetimes to the future of the collision on either side of the domain wall. The metric in each region is of the form: 
\begin{equation}\label{metric}
ds_{o,C}^{2} = - a_{o,C}(z)^{-1} dz^2 +  a_{o,C}(z) dx^2 + z^2 dH_{2}^{2}
\end{equation}
where 
\begin{equation}
dH_{2}^{2} = d \chi^2 + \sinh^2 \chi d\phi^2
\end{equation}
is the metric of a spacelike 2-hyperboloid. For a general vacuum solution we have
\begin{equation}
\label{eq-aoc}
a_{o,C} = 1 - \frac{2 M_{o,C}}{z} + H^2_{o,C} z^2,
\end{equation}
where the subscripts $o,C$ specify the metric on the side of the observation bubble and colliding bubble respectively. The gravitational back reaction of the null shells gives rise to mass parameters $M_o$ and $M_C$ in the region to the future of the collision. Setting $M_{o,C}=0$ yields de Sitter (for $H^2 > 0$) or Anti de Sitter space (for $H^2 < 0$) in an explicitly hyperbolic form of the coordinates. 

The energy-momentum tensor on the post-collision domain wall is assumed to be
\begin{equation}
\label{eq-ktuv}
T_{a b} = -\sigma_{oC} \delta (z-z_{\rm wall}) \gamma_{a b},
\end{equation}
where $a,b = 0 ,1, 2$ and $\gamma_{ab}$ is the metric on the worldsheet of the wall. The first junction condition is to require that the $z$ coordinate (which represents the physical radius of curvature of the 2-hyperbola labeled by $(x,z)$) is continuous across the wall; the $x$ coordinate is necessarily discontinuous. Integrating Einstein's equations across the wall then relates the jump in the extrinsic curvature $K_{ab}$ to the tension of the wall $\sigma_{oC}$: 
\begin{equation}
K_{C \ b}^{a} - K_{o \ b}^{a} = - 4 \pi \sigma_{oC} \delta_{b}^{a}.
 \end{equation}
This yields a set of equations of motion for the location $z(\tau)$, $x(\tau)$ of the wall, where $\tau$ is the proper time in a frame riding with the wall. This parametric set of variables can then be manipulated to obtain $z(x)$. The relevant equations of motion and their solutions can be found in AJ and CKL1.

For the null shells, the energy momentum tensor is assumed to be
\begin{equation}
T_{\mu \nu} = \sigma_r l_{\mu} l_{\nu} \delta ({\bf x}_{shell}),
\end{equation}
where $l_{\mu}$ is a vector tangent to the null shell and $\sigma_r$ is the energy density on the shell. Integrating Einstein's equations across the wall gives a junction condition of
\begin{equation}
K_1 - K_2 = 8 \pi \sigma_r,
\end{equation}
where $K$ is the trace of the extrinsic curvature, and we have chosen a convention such that the shell is moving from region `1' into region `2.' This junction condition relates the mass parameter in the region to the future of the null shell to the surface energy density:
\begin{equation}\label{energydensity}
M_{o,C} = 4 \pi \sigma_r z^2.
\end{equation}

Finally, we must impose energy and momentum conservation at the location of the collision. The boost angle between the rest frames of two colliding walls (each labeled by the 4-velocity $U^\mu$ of an observer in that rest frame) can be defined by
\begin{equation}
\label{eq-boostcosh}
g_{\mu \nu} U_{i}^{\mu}  U_{j}^{\nu} = \cosh{\pm \xi_{ij}}
\end{equation}
where $g_{\mu \nu}$ is the metric in the spacetime between the colliding walls, and where the angle is negative when one wall is incoming (one of the initial state, pre-collision, walls) and the other is outgoing (one of the final state, post-collision, walls) and positive when both are incoming or outgoing. Performing a series of boosts between an arbitrary number of domain walls, we should find that upon coming back to the original frame, the total sum of the boost angles is zero:
\begin{equation}\label{conservation}
\sum_{i}^{n} \xi_{i\ i+1} = 0.
\end{equation}
This turns out to be equivalent to imposing energy and momentum conservation~\cite{Langlois:2001uq,Freivogel:2007fx}.

\begin{figure*}[htb]
\includegraphics[width=13 cm]{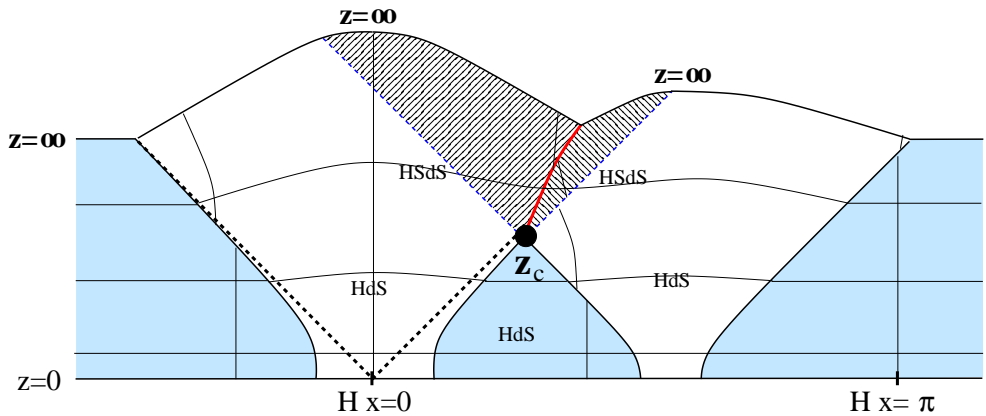}
\caption{The collision between two de Sitter bubbles displayed in the $(x,z)$ plane. Lines of constant $z$ and $x$ are drawn: note that $z$ is continuous across each of the junctions, but $x$ is discontinuous as required by the junction conditions. The post-collision domain wall is drawn as the solid red line, and the null shells as dashed blue lines. The causal future of the collision (shaded on the diagram) is Hyperbolic SdS, with a different mass parameter on each side of the post-collision domain wall.
  \label{fig-colldiag}
}
\end{figure*}

Putting the various pieces together, we can construct the full collision spacetime. An example is shown in Fig.~\ref{fig-colldiag}, which depicts a conformal slice of two colliding dS bubbles. In the collision frame, the separation between two bubbles' nucleation sites fully specifies the kinematics. This can be parametrized by the value of $z$ at which the bubbles first collide, $z_c$. There are then the parameters $\{H_F, H_c, H_o, \sigma_{oC} \}$ that are determined by the underlying potential, and $\{ M_o, M_C\}$ which are introduced to conserve energy and momentum at the collision.

Summarizing the important results obtained from this framework:
\begin{itemize}
\item The asymptotic behavior of a post-collision domain wall is determined completely by the vacuum properties of the two colliding bubbles $H_{C}$ and $H_o$ together with the tension of the post-collision domain wall $\sigma_{oC}$. The post-collision domain wall has constant acceleration unless $16\pi^2 \sigma_{oC}^2 = - (H_o - H_C)^2$ (the BPS condition of Ref.~\cite{Freivogel:2007fx}), in which case it is timelike. When 
\begin{equation}\label{eq:domainwalloutcond}
H_{C}^2 - H_{o}^2 + 16 \pi^2 \sigma_{oC}^2 > 0
\end{equation}
the post collision domain wall asymptotically moves away from the observation bubble interior; when
\begin{equation}
H_{C}^2 - H_{o}^2 + 16 \pi^2 \sigma_{oC}^2 < 0
\end{equation}
it moves towards the observation bubble interior. 
\item In general, it is only possible to derive one relation between $M_o$ and $M_C$ using Eq.~\ref{conservation}, and so their individual values cannot be precisely determined without a more detailed model of the microphysics of the collision. However, in the limit where the initial radii of the colliding bubbles are small and the post-collision domain wall has small tension compared with $H_{o}$ and $H_{c}$, it is possible to derive a relation~\cite{Aguirre:2007wm} fixing $M_{o}$ to be:
\begin{equation}\label{Mo}
M_o \simeq \frac{(H_F^2 - H_o^2) (1+H_o^2 z_c^2)}{2 (1+H_F^2 z_c^2)} z_c^3 m_p^2.
\end{equation}
This coincides with the result for the collision between identical dS bubbles from CKL1 and with the result for `mild' collisions in AJ.
\item Depending on the kinematics, the motion of the post-collision domain wall can possess a turning point in $x$. That is, the domain wall can go from incoming to outgoing or vice versa. This occurs when the quantity
\begin{equation}\label{zbeta}
z_{\beta_{o}} \equiv \left[\frac{2 (M_C - M_o) }{ H_C^2 - H_o^2 + 16 \pi^2 \sigma_{oC}^2 }\right]^{1/3},
\end{equation}
is real and exceeds $z_c$.
\end{itemize}

\subsection{Numerical and semi-analytic results}\label{sec:numericalsims}

While the gross features of the collision spacetime are captured quite well by the junction condition formalism described in the previous section, it leaves open a number of important questions. Perhaps the most pressing concerns how the collision affects the surfaces of constant density inside of the bubble. This determines what kind of cosmology can exist to the future of a collision, and therefore what an observer in this region might see. To perform the detailed analysis of the bubble interior, it is necessary to employ numerical~\cite{Hawking:1982ga,Kosowsky:1991ua,Blanco-Pillado:2003hq,Aguirre:2008wy,Easther:2009ft} and semi-analytic~\cite{Chang:2008gj} models of the collision spacetime. Using these tools, the following questions relevant to the structure of the collision spacetime have been addressed in the literature
\begin{itemize}
\item Can spacelike surfaces of homogeneity inside of the observation bubble exist to the future of a collision?
\item Can the effects of the collision be mild enough to be consistent with the observed level of homogeneity and isotropy?
\item Are there predictions for the generic form of anisotropies and/or inhomogeneities produced by a bubble collision?
\item In what cases can an epoch of inflation occur to the future of a collision, therefore diluting curvature and seeding structure?
\end{itemize}
We now describe some of the existing results and directions for future progress in this area, focusing on the analysis performed in AJT, CKL2, and Easther et. al.~\cite{Easther:2009ft}.

\subsubsection{Numerical simulations in 1+1D}
As in the thin-wall matching problem, the SO(2,1) hyperbolic symmetry of the collision region between two bubbles allows for a $1+1$-dimensional description of the full collision spacetime. The original simulations of Hawking, Moss, and Stewart~\cite{Hawking:1982ga} and the subsequent analysis in AJT focused on the simulation of bubble collisions in flat space, neglecting gravitational effects (a computation of bubble collisions including gravity but in a different context was given in~\cite{Blanco-Pillado:2003hq}; a background Hubble expansion was included by Easther et. al.). This is generally a good approximation as long as the energy densities do not become too large, and the simulation region is sufficiently small to neglect the expansion of the background false-vacuum. This requires that the nucleation radii of the bubbles, and their initial separation, are much less than the false vacuum Hubble size $H_F^{-1}$. 

\begin{figure}
\includegraphics[width=7.5cm]{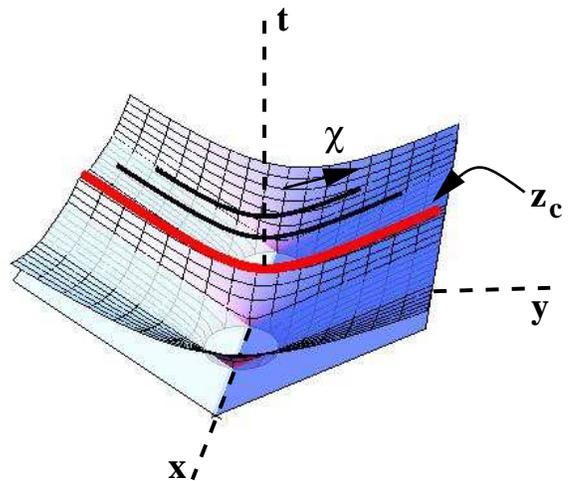}
\caption{
The collision of two bubbles in Minkowski space with nucleation centers located on the $x-$axis. The hyperbolic coordinates Eq.~\ref{eq-HyperbolicCoordTransform} cover the portion of the spacetime above the planes, and the collision surface, outlined by the thick red line, is located at a constant hyperbolic position $z_c$. Moving along the constant $z$ hyperboloids corresponds to increasing $\chi$ in Eq.~\ref{eq-xzmetric}.
  \label{fig-minkcoll}
}
\end{figure}

In this case, we can define the coordinates
\begin{eqnarray}  \label{eq-HyperbolicCoordTransform}
&& t = z \cosh \chi,\qquad\quad\,  x = x, \\ \nonumber
&& y = z \sinh \chi \cos \varphi, \ \  w = z \sinh \chi \sin \varphi,
\end{eqnarray}
in terms of which the flat-space metric is
\begin{equation}\label{eq-xzmetric}
ds^2 = -dz^2 + dx^2 + z^2 (d \chi^2 + \sinh^2 \chi d\varphi^2),
\end{equation}
and $\phi(z,x)$ obeys
\begin{equation}\label{eq-hypevolve}
\partial_z^2\phi-\partial_x^2\phi+{2\over z}\,\partial_z\phi
  = -{dV\over d\phi}.
\end{equation}
Each point in the $(z,x)$ simulation plane corresponds to a two-hyperbola of radius $z$, as depicted in Fig.~\ref{fig-minkcoll}.

\begin{figure*}
\includegraphics[width=12cm]{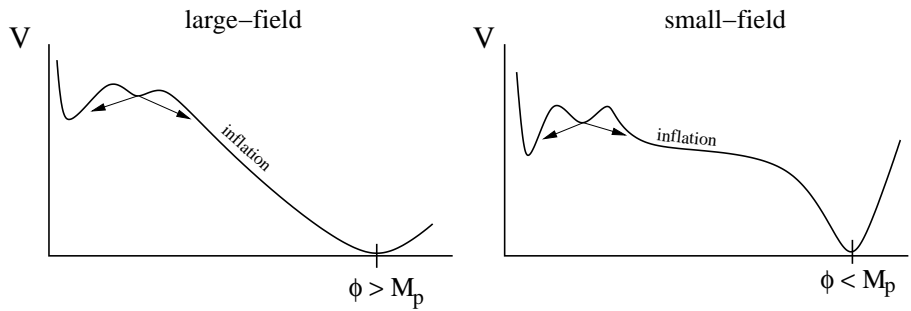}
\caption{The two classes of potential studied in AJT. The false vacuum in each case can decay into another high energy vacuum or to an inflating region of the potential. Potentials of the type shown on the left yield large-field models of inflation, in which the field rolls over a super-planckian distance. Potentials of the type shown on the right yield small-field models of inflation, where the field loiters near a critical point of the potential during inflation (this is also referred to as `Accidental inflation'~\cite{Linde:2007jn}.)
  \label{fig-twopotentials}
}
\end{figure*}

In AJT, two classes of single-field potentials $V(\phi)$ were studied, both of which could drive a phenomenologically viable epoch of open inflation if gravitational effects were included. These are depicted in Fig.~\ref{fig-twopotentials}.  In each case the false vacuum has two decay channels, one to a high energy vacuum (inside the collision bubble), and the other to an inflating region of the potential (inside the observation bubble). The two classes of potential differ in the properties of the inflationary region, which is of the `large-field' (left panel of Fig.~\ref{fig-twopotentials}) or `small-field' (right panel of Fig.~\ref{fig-twopotentials}) type. Most single-field models of inflation can be categorized as one of these two types.

To set up the initial conditions for collisions, domain wall configurations must be found that interpolate across the barriers separating the false vacuum from the two other minima. These domain walls can be constructed numerically from the Euclidean field equations
\begin{equation}\label{eq-CDLinitialdata}
\partial_{\tau_E}^2 \phi + \frac{3}{\tau_E} \partial_{\tau_E} \phi= \frac{dV}{d\phi} \,,
\end{equation}
with $\tau_E \equiv \sqrt{r^2+t^2}$~\cite{Coleman:1977py}. One initial condition $\phi(\tau_E = 0)$ is in the basin of attraction of the true vacuum (where $\partial_{\tau_E} \phi = 0$), and we require that the false vacuum is reached as $\tau_E \rightarrow \infty$. The initial field configuration on the simulation grid at $t=0$ is then generated by identifying $\tau_E=r$.

The specific potentials and initial conditions implemented in the numerical simulations were chosen to minimize the importance of gravitational effects in the simulation region. In particular, the initial bubble size and separation were smaller than the false vacuum Hubble scale, and the energy densities were everywhere much less than planckian. Scales only enter into the problem to assure that phenomenologically viable models of inflation could be generated, and so the qualitative results should hold in a more realistic model.

\begin{figure}
\includegraphics[width=7.5cm]{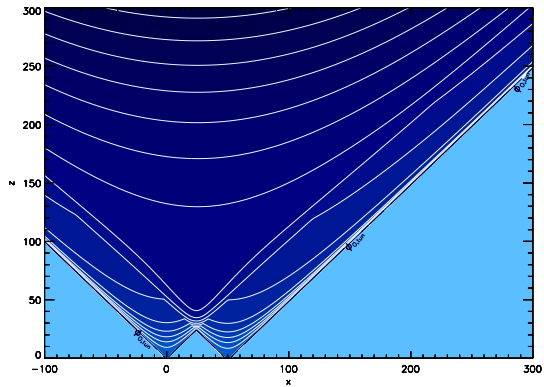}
\caption{The simulated collision of two identical observation bubbles generated from a large-field type potential (sketched in the left panel of Fig.~\ref{fig-twopotentials}). To the future of the collision, the surfaces of constant-field are well fit by hyperbolas, implying that a boost symmetry of the field configuration is spontaneously generated. The field remains in the inflationary region of the potential after the collision. 
  \label{fig:twosame}
}
\end{figure}

\begin{figure}
\includegraphics[width=7.5cm]{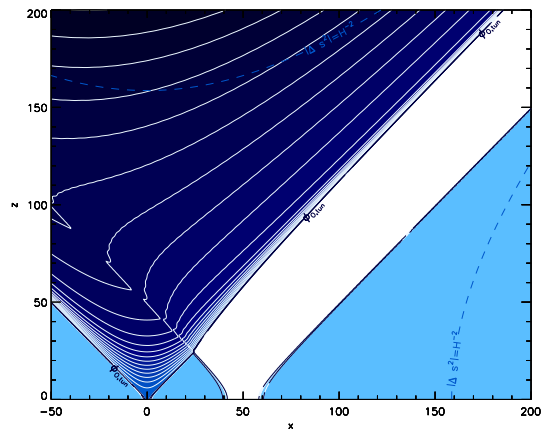}
\includegraphics[width=7.5cm]{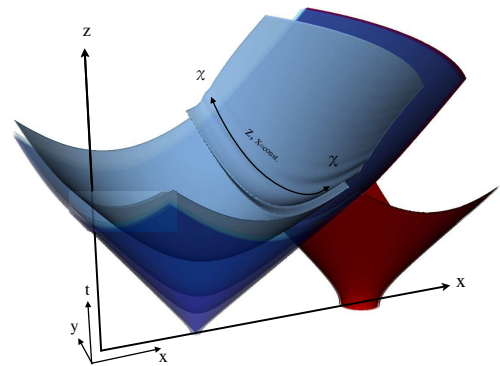}
\caption{{\em Top:} The collision between an observation bubble and a collision bubbble generated from a large-field type potential. A post-collision domain wall forms, which accelerates away from the interior of the observation bubble. To the future of the collision, surfaces of constant field again approach spacelike surfaces of homogeneity, and are well fit by hyperbolas.  {\em Bottom:} 2+1 dimensional visualization of the same collision.
  \label{fig:twodiff}
}
\end{figure}

\begin{figure}
\includegraphics[width=7.5cm]{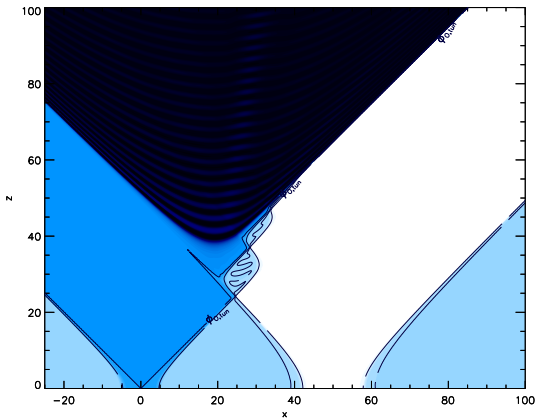}
\caption{The collision between an observation bubble and a collision bubble generated from a small-field type potential. To the future of the collision inside of the observation bubble, the field is pushed away from the inflating region of the potential to the low-energy minimum. Inflation does not occur to the future of the collision. 
  \label{fig:accident}
}
\end{figure}

The main conclusions of AJT can be understood from the simulations shown in Figures~\ref{fig:twosame},~\ref{fig:twodiff}, and~\ref{fig:accident}.
\begin{itemize}
\item Fig.~\ref{fig:twosame} shows the collision between two identical observation bubbles generated from the large-field potential. Spacelike surfaces of constant field form to the future of the collision. The field remains on the inflationary region of the potential, allowing for a phenomenologically viable cosmology to arise in the future of the collision. Studying the surfaces of constant field in more detail, AJT found that at late times they fit to hyperbolas with high accuracy. Thus, a boost symmetry along the direction separating the two colliding bubbles is {\em spontaneously} generated at late times in the post-collision region. However, the origin of this set of hyperbolas does not match the origin of the 2-hyperbolas ($z=0$) defined by each point in the simulation region, and the surfaces of constant field therefore have an intrinsic anisotropy (though this becomes negligible far from the collision).
\item Fig.~\ref{fig:twodiff} depicts the collision between an observation and colliding bubble generated from the large-field potential. After the collision, a domain wall forms separating the interiors of the two bubbles, which then accelerates into the colliding bubble. Just as in the thin-wall matching analysis, this occurs because the energy density on the inflating region of the potential inside the observation bubble is of lower energy than the vacuum phase inside of the colliding bubble. From the collision, a  null disturbance propagates into the observation bubble, to the future of which the field is advanced slightly in its motion along the inflationary part of the potential. As for the collision of two identical bubbles, spacelike surfaces of homogeneity form to the future of the collision, on which the field is in the inflationary region of the potential. Further, sufficiently far up the accelerating post-collision domain wall, the surfaces of constant field are hyperbolas, implying that there are in fact infinite spacelike surfaces of homogeneity to the future of the collision. The appearance of these hyperbolic surfaces can be viewed as a consequence of the constant acceleration of the post-collision domain wall.
\item In Fig.~\ref{fig:accident}, the collision between an observation and colliding bubble generated from the small-field potential is shown. To the future of the collision, the field inside of the observation bubble is pushed away from the `loitering' region of the potential. Inflation does not occur to the future of the collision in this model, rendering it `falsifiable' (or perhaps `fatal') as defined in the Introduction. AJT therefore concluded that inflation can generically occur to the future of a collision only if the potential is of the large-field type.
\end{itemize}

\subsubsection{Numerical simulations in 3+1D}
\begin{figure}
\includegraphics[width=7.5cm]{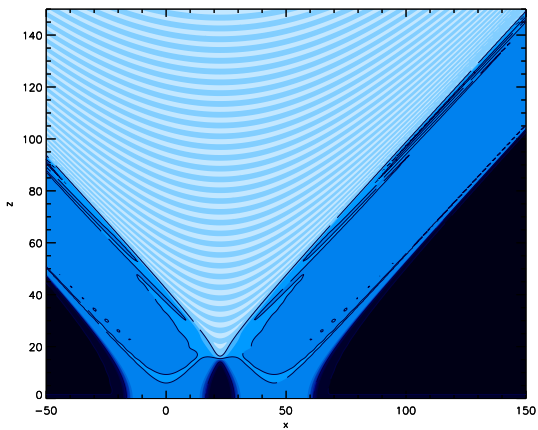}
\caption{Collision between two identical bubbles of the intermediate vacuum in the triple-well potential of Ref.\cite{Easther:2009ft}; the energy released in the collision forms a region of the lowest vacuum energy.
  \label{fig-eastherreprod}
}
\end{figure}

More recently, bubble collisions were studied numerically in Ref.\cite{Easther:2009ft} using a four-dimenional code including background Hubble expansion. In the collisions studied by these authors, the SO(2,1) hyperbolic symmetry was shown to be preserved to good accuracy, justifying the assumptions made in previous studies. As an application of the code, an underlying potential with three vacua of decreasing energy was studied. Bubbles containing the intermediate vacuum underwent collisions in a background of the highest energy vacuum. Interestingly, the gradient energy contained in the colliding walls is generically large enough to form a pocket of the lowest energy vacuum in the spacetime region to the future of the collision. This pocket expands due to the outward pressure gradient, and in this sense collisions can be thought of as a classical mechanism for the nucleation of a vacuum `bubble.' The kick experienced by the field is similar to the dynamics responsible for ending inflation to the future of a collision in the small field models studied by AJT.  This calculation is reproduced (in 1+1D) in Figure~\ref{fig-eastherreprod}.

\subsubsection{Semi-analytic analysis}
An alternative study of the behavior of inflation to the future of a collision was performed in CKL2. The authors analyzed the behavior of a test-field (meant to describe the inflaton) in the background geometry of the thin-wall collision spacetime. The test-field was fixed to be constant along the post-collision domain wall, and to evolve along the open-slicing surfaces of constant $\tau$ outside of the future light cone of the collision. The field is matched across a null wall following this lightcone, and this forces the surfaces of constant field to be either advanced or retarded from those outside the region influenced by the collision, depending on the underlying potential landscape.  This leads to a slightly different number of e-folds of inflation in different regions of the bubble. In this calculation, where spacetime expansion is taken into account, it can again be confirmed (see AJ) that infinite spacelike surfaces of homogeneity develop to the future of the collision at sufficiently late times. We therefore expect that the habitable volume to the future of at least some types of collisions can be infinite.

\subsubsection{Future directions}
Currently, numerical analyses of bubble collisions reveal a number of important qualitative results (e.g. that observers can exist to the future of at least some collision types), but do not provide a detailed quantitative foundation upon which to base conclusions about the observability of `detectable' or `falsifiable' collisions. Some promising extensions for future work include:
\begin{itemize}
\item Simulations including gravitational effects, in which the metric describing the bubble interior after a collision could be solved for precisely. Using such a model, it is plausible that one could make quantitative predictions for the deviations from homogeneity and isotropy in the future of a collision. 
\item Multi-field potentials. Depending on the assumed couplings, it is possible that colliding bubbles do not even interact.
\item Including interaction with radiation and other fields. Again, depending on the assumed underlying theory, colliding bubbles might produce a variety of debris, and interact in different ways.  
\end{itemize}

\section{Observable signatures}\label{sec:obssignatures}

Although the numerical and analytic models above do not give a precise and detailed understanding of the structure of bubble collisions, they do afford sufficient qualitative insights that we can outline a number of potentially observable signatures of bubble collisions.  Since many of these are only worked out roughly in the literature, we will aim primarily to categorize and summarize these possibilities, largely as a guide toward future work.

\subsection{General considerations: symmetries and angular scales}\label{sec:angularscale}

An observer's ``sky" is defined by the intersection of their past light cone with a particular (or perhaps a set of) equal-time slices in the collision spacetime. Because the azimuthal symmetry present in the metric of both colliding bubbles remains unbroken, a single collision retains an azimuthal symmetry on the sky.  For example, the intersection of a null shell of collision debris (which denotes the causal boundary of affect of the collision) with a surface of constant $\tau$ appears as circle on the observer's sky, the disk interior to which represents the part of this surface possibly affected by the collision.  While some effects (mentioned below) might act to break this symmetry, they are expected to be subdominant, so this is a clear signature to seek in the CMB or other observables~\cite{Aguirre:2007gy,Chang:2008gj}.

The `disk of influence' of a given collision subtends some angular scale on the sky, which can be tied to the details of the collision. In terms of the CMB, this scale can be defined by the triple intersection of an observer's past light cone, the last scattering surface (LSS), and the boundary of the region inside of the bubble affected by the collision (as defined for example by the incoming shell of collision debris). This problem is in general difficult, but can be worked out in the approximation that (a) the collision does not affect the bubble interior (allowing us to use the original open FLRW foliation to define constant density hypersurfaces), and (b) that the collision debris propagate into the bubble along null rays. 

Within this setup, the angular scale at very early times ($\tau \rightarrow 0$) was first calculated in AJS. However, a quantity more relevant for observation is the angular scale on the LSS, which can be quite different as detailed by FKNS. The angular scale depends upon the nucleation position of the incoming bubble, the position of the observer, and the cosmology inside the observation bubble (described in Sec.~\ref{sec-cosmobubble}). For an observer at the origin (as we discuss in Sec.~\ref{sec-probabilities}, when the original FRW foliation is retained, it is always possible to translate the observer to the origin), in Appendix~\ref{sec:obsang} we derive an exact expression for the observed angular scale $\psi$ on the reheating surface, given by
\begin{widetext}
\begin{equation}
\label{eq:costhetadS}
\cos \left( \frac{\psi}{2} \right) = \frac{1}{\sinh (H_I \tau_{\rm reh}) \sinh \xi_{\rm reh}} \left[ \frac{\left( 1 + H_I^2 z_c^2 \right) - \left( 1 - H_I^2 z_c^2 \right) \cosh (H_I \tau_{\rm reh})}{2 H_I z_c \gamma} + v \sinh (H_I \tau_{\rm reh}) \cosh \xi_{\rm reh} \right],
\end{equation}
\end{widetext}
where $H_I$ is the Hubble scale during inflation, $\xi_{\rm reh}$ is the radial distance out to which the observer can see on the reheating surface, $\tau_{\rm reh}$ is the open slicing time at reheating, $z_c$ is the hyperbolic radius at which the bubbles first collide in a frame where they nucleate simultaneously (the `collision' frame as defined in the simulations of the previous section, and discussed below in Sec.~\ref{sec-refframes}), and $v$ is the relative velocity between the rest frame of the collision and the rest frame of the observer. Because $\Omega_c$ is initially very small after inflation, and grows little between last scattering and reheating, $\xi_{\rm reh} \simeq \xi_{\rm ls}$ to good accuracy (see Eq.~\ref{eq-conftime}). The output from Eq.~\ref{eq:costhetadS} should therefore be equally valid for the observed angular scale at reheating and last scattering, but will be altered for general $\xi_{\rm view}$. 

\begin{figure*}
\includegraphics[width=8cm]{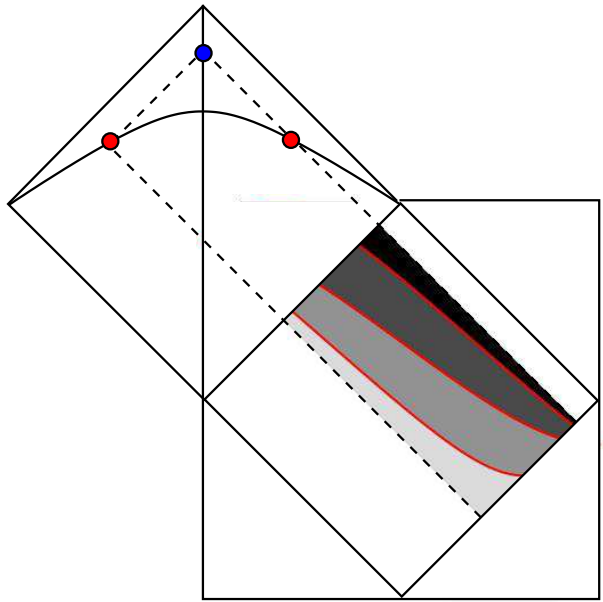}
\caption{A map between the nucleation point of a bubble and the observed angular scale $\psi$ in the small-bubble approximation. Here, we consider the limit where $H_{I} \ll H_{F}$, $N_e \gg 1$, and $\xi_{\rm reh} \sim 1$ (we choose such an unrealistic value to display the map more clearly; for a consistent value $\xi_{\rm reh} \sim 0.18$, the map is compressed). The contours in the false vacuum correspond to an observed angular scale (from dark to light) of $\pi/2, \pi, 3\pi/2$.
\label{fig-anglesmap}}
\end{figure*}

In a frame in which the observer is at the origin (the `observation' frame defined in Sec.~\ref{sec-refframes}), a bubble nucleated at each point in the false vacuum maps uniquely to a given observed angular scale as shown in Fig.~\ref{fig-anglesmap}. The observed angular scale of the collision boundary can range from $0 < \psi < 2 \pi$ for different collision events. Because an observer has access to only a finite portion of an earlier surface of constant $\tau$, it is possible to be to the future of a collision without seeing the null shell of debris. In this case, the effects of the collision are spread out in an azimuthally symmetric pattern over the observer's entire sky.

\subsubsection{Symmetry breaking effects}

The SO(2,1) symmetry (including the azimuthal symmetry) of the collision of two vacuum bubbles is a consequence of the SO(3,1) symmetry of each of the colliding bubbles. There are a number of plausible situations where this assumption can be violated:
\begin{itemize}
\item Clearly, more than two bubbles colliding will reduce the symmetry, and violate axisymmetry of perturbations on, e.g., the CMB in the obvious way.
\item If the bubbles do not evolve in vacuum, the boost symmetry of each is broken and the bubbles can have a different structure (see~\cite{Garriga:1997ht,GarciaBellido:1998wd} and the recent paper~\cite{Fischler:2007sz}). 
This effect can occur whenever the false vacuum is evolving, but should be small to the extent that nucleation rates are small (so that the false vacuum is `old' for any given bubble).
\item Similarly, fluctuations in the false vacuum -- while being statistically SO(3,1)-invariant -- lead to violations of the SO(2,1) symmetry of the collision spacetime in any given realization. This is similar to what occurs in Open Inflation, where such fluctuations break the SO(3,1) symmetry of the undisturbed bubble~\cite{GarciaBellido:1997uh}. These effects should be subdominant to bubble collisions to the extent that the fluctuations  are small in amplitude.
\item Fluctuations in the colliding bubble walls, and in the post-collision domain wall would be present because vacuum bubble walls are unstable to such perturbations growing (see e.g.,~\cite{Adams:1989su}, and~\cite{Aguirre:2005xs} for the case of bubbles without a boost symmetry). This would appear to be a small effect, as for any accelerating wall the growth of perturbations tends to be swamped by the growth in the radius of curvature of the wall. However, perturbations will be important in cases where a collapsing wall is produced. The false vacuum pockets found in simulations of bubble collisions~\cite{Hawking:1982ga} provides one example.  
\end{itemize}

\subsection{Mechanisms for affecting cosmological observables}

\subsubsection{The shape of the reheating surface}

\begin{figure*}
\includegraphics[width=16cm]{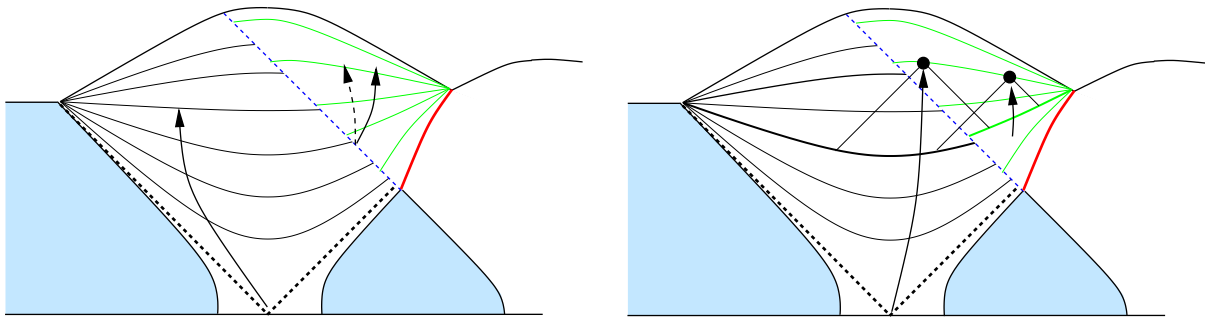}
\caption{A model for the structure of post-collision regions.  In the left panel, lines of constant field are superimposed on the post-collision spacetime. Outside the future light cone of the collision, observers that are comoving with respect to the original open FLRW foliation follow the solid worldline. Inside the future light cone of the collision, the surfaces of constant field are disturbed, and comoving observers with respect to this new foliation would follow the dashed worldline. Note that there is a relative boost between these two foliations at any point. In the right panel, we show two observers: a `native born' observer associated with structure formed in the undisturbed part of the bubble, and a `foreign born' observer associated with structure formed in the region of the bubble affected by the collision. These observers both have causal access to both affected and unaffected portions of the  surface of last scattering (dark surface of constant field). Each observer is boosted with respect to some portion of the surface of last scattering, the distance to which then becomes angle-dependent.
\label{fig-collspacetime}}
\end{figure*}

As discussed in  Sec.~\ref{sec-genericcollision}, a collision can significantly alter the structure of the observation bubble interior, breaking the SO(3,1) symmetry of the original undisturbed open FLRW universe. The altered shape of the reheating surface, and later surfaces such as the LSS that it evolves into, has potentially observable effects on the CMB. One such effect was examined in some detail in CKL2. The setup is shown in the conformal slice Fig.~\ref{fig-collspacetime}. The surfaces of constant field (and hence the natural foliations) inside and outside the future light cone of a collision are not synchronized, as indicated by the two comoving trajectories to the future of the collision in the left panel of Fig.~\ref{fig-collspacetime}. The observer might be moving along a comoving worldline of either foliation, depending upon whether the observer is associated with a structure that formed entirely to the future of the collision (a `foreign-born observer'), or formed outside of the collision region and entered it at late times (a `native born' observer).

If the collision is sufficiently disruptive to structure formation, this singles out the native-born observers. This case is treated by CKL2. These authors used the semi-analytic formalism described in Section~\ref{sec:numericalsims} to obtain the modified reheating surface. Assuming a cosmology consisting of an epoch of inflation followed by a period of radiation domination, they then calculate the redshift of photons coming to the observer from the LSS as a function of angle (assuming the redshift to be of the same functional form as the redshift from the reheating surface). 
 
In calculating the redshift, there are two relevant effects. First, the distance to the LSS gains angular dependence; this can be seen in the right cell of Fig.~\ref{fig-collspacetime}, where the reheating surface is closer in the direction of the domain wall. In addition, there is a doppler effect due to the relative boost between the comoving foliation outside (which the observer follows) and inside the collision region. Depending on the details of the model, and the position of the observer, the difference in redshift can be appreciable. 

This redshift profile can then be translated into a temperature profile in the CMB sky:
\begin{equation}
T(\mathbf{\hat{n}}) = T_0' r(\mathbf{\hat{n}}) \left[ 1+ \delta (\mathbf{\hat{n}}) \right],
\end{equation}
where $T_0'$ sets the overall temperature, $r(\mathbf{\hat{n}})$ is the redshift to the surface of last scattering as a function of the viewing angle, and $\delta (\mathbf{\hat{n}})$ is a gaussian random variable encoding the primordial temperature anisotropies. The altered reheating surface in this model therefore merely acts as a background modulation on the fluctuations $\delta (\mathbf{\hat{n}})$ in the undisturbed region of the bubble. The results of this analysis include:
\begin{itemize}
\item Centering the collision on $\theta = 0$, because of the azimuthal symmetry of the collision, any excess power is in the $m=0$ modes. 

\item A power asymmetry arises due to the greater amplitude of temperature fluctuations in directions with a higher background temperature.

\item CKL2 found that in their model the fractional change in redshift increases or decreases roughly linearly with $\cos \theta$, where the direction of the collision is assumed to be along $\theta = 0$. 

\item The maximum value of the fractional redshift or blueshift due to the collision is observationally bounded by the observed CMB. As a fraction of the redshift to the undisturbed portion of the reheating surface, CKL2 quote a limit on the order of $|\frac{r_{\rm out}}{r_{\rm in}} - 1| \alt 10^{-3}$. 

\item In the power spectrum, those $C_{l}$ with $l$ of order the angular size of the collision disk (and its harmonics) are boosted. Collisions that appear as larger discs on the sky of the observer therefore primarily affect low-$l$ multipoles, while collisions with a small angular scale primarily affect higher $l$ multipole moments.

\end{itemize}

This analysis provides a good guide to what types of observable effects may arise due to a bubble collision, and future work should explore different assumptions and expand upon a number of additional interesting effects. If structure can form to the future of a collision, then one can consider the effects seen by foreign-born observers. In addition, since native- and foreign-born structures will follow different comoving congruences, there may be bulk flows visible at the scale of galaxies or clusters of galaxies~\cite{Chang:2008gj}.  Potentially observable non-gaussianities may also be sourced by the collision. Including more details in the model, it is important to study the evolution of primordial fluctuations in the perturbed background of the collision spacetime, which could lead to a number of other qualitatively different effects. The presence of correlated signatures would be exciting evidence that a specific feature in the CMB could correspond to a bubble collision, and a more detailed quantitative study of the strength of such correlations is certainly merited. 

\subsubsection{Observing the collision debris}\label{sec:observingdebris}

Within the approximations of the thin-wall collision problem of Sec.~\ref{sec:thinwallcollisions}, another potentially important effect is the energy density associated with the null shell of collision debris. The energy density $\sigma_{N_e}$ in this shell after $N_e$ e-folds of inflation can be obtained from Eq.~\ref{energydensity} by finding the coordinate $z$ at which the shell trajectory intersects the reheating surface. This calculation is shown in Appendix~\ref{sec-nullshell} to give a result (Eqs.~\ref{eq:sigmaintermediate},~\ref{eq:sigmaapprox}) in terms of collision coordinate $z_c$. 

For `mild collisions' (where Eq.~\ref{Mo}) applies), typical values $z_c \sim H_F^{-1}$ (see Sec.~\ref{sec-zcdist} below), and assuming $H_F$ is somewhat larger than $H_I$ (the Hubble scale during inflation), this gives
\begin{equation}
\frac{\sigma_{N_e}}{H_I^3} \simeq \frac{1}{ 4 \pi \cosh^2 N_e } \frac{m_p^2}{H_F^2} \frac{H_I}{H_F}.
\end{equation}
For high scale inflation, the energy density in the radiation shell at reheating will be negligible. If $H_F$ and $H_I$ are set approximately by the GUT scale, then the energy density in the shell is diluted after $\sim 15$ e-folds of inflation, far fewer than the required $N_e \sim 60$. However, if $H_F$ and $H_I$ are set by a lower energy scale, then the energy density might be non-negligible at reheating. In the extreme case, where $H_F$ and $H_I$ are set by TeV scale, then even after $70$ e-folds (many more than the required 30 or so, see Sec.~\ref{sec-cosmobubble}), the energy density in the shell will be greater than $H_I^3$. Depending on how the debris interact with standard model degrees of freedom, this could cause local changes in the properties of reheating.

The energy density in the radiation shell can be enhanced if $z_c$ is much larger than $H_F$, although it will be very rare for a comoving observer to encounter such a collision (the distribution for $z_c$ is discussed in Sec.~\ref{sec-probabilities}). In the limit where $z_c \gg H_{I,F}^{-1}$,
\begin{equation}
\frac{\sigma_{N_e}}{H_I^3} \simeq \frac{1}{ 2 \pi \cosh^2 N_e } \frac{m_p^2}{H_I^2} H_I z_c
\end{equation}
In the case of GUT scale inflation, then it is necessary to have $z_c \sim 10^{40} H_I^{-1}$ for $\frac{\sigma_{N_e}}{H_I^3} \agt 1$ after $60-70$ e-folds of inflation. However, this is predicated on the small-bubble approximation being satisfied. If it is not, the mass parameter appearing in Eq.~\ref{eq:sigmaintermediate} will saturate at some $z_c$, leading to a maximum enhancement of $\sigma_{N_e}$.

The radiation shell also causes  a discontinuity in the derivatives of the metric. Photons crossing the shell are red-shifted, possibly giving rise to a cold spot inside of the disc defining the collision boundary. As discussed in Appendix~\ref{sec-nullshell}, this effect always appears to be small (exponentially suppressed by the number of e-folds). However, the metric discontinuity should be much larger at the beginning of inflation.  Fluctuations produced during inflation can be strongly redshifted as they cross the shell of collision debris, potentially leading to observable effects on large scales in the CMB.

\subsubsection{Changing the cosmological history to the future of the collision}

It is quite possible that the properties of inflation to the future of the collision are subtly or even dramatically different than outside the collision region. The array of such possibilities is very large and should be explored systematically. Here, we give a brief sampling: 
\begin{itemize}
\item The collision injects energy into the inflaton field, leading to a different number of e-folds in different regions (and a distorted shape of the reheating surface, as discussed above). It is even possible, as we saw for the small-field models in Sec.~\ref{sec:numericalsims}, that inflation does not occur at all to the future of a collision or that pockets of other vacua are generated~\cite{Easther:2009ft}. 
\item Another possibility is that bubble collisions could seed topological inflation inside of inflating domain walls separating the two colliding bubbles.
\item In a multi-field model, a collision could cause the field trajectory to be different in some regions of spacetime, probably leading to a rather different spectrum of fluctuations and set of reheating properties in various observationally accessible regions. 
\item In the case where the fields forming a pair of colliding bubbles are completely uncoupled, there would be neither collision debris nor post collision domain wall: the two bubbles would simply overlap~\cite{BlancoPillado:2009di}. However, unavoidable gravitational interaction would lead to anisotropies in the overlap region due to the presence of some time-dependent `dark' energy density. This might, for example, occur in the collision of bubbles arising from flux compactifications~\cite{BlancoPillado:2009di}. 
\end{itemize}

\section{Bubbles and collision probabilities in an eternally inflating background}
\label{sec-probabilities}

As described in Secs.~\ref{sec-genericcollision} and~\ref{sec:obssignatures}, the effect of a bubble collision on a putative observer can be Fatal, Falsifiable, Detectable, Invisible {\em or} Nonexistent, depending upon the collision parameters and inflaton potential. The possibility of Falsifiable or Detectable bubbles is very interesting, but only insofar as observers seeing them are relatively common as compared to those seeing only Invisible or Non-existent bubbles.

Assessing these relative probabilities entails defining a measure over all of the possible types of observers inside the full set of bubbles. Such measures in eternal inflation are fraught with difficulties and ambiguities (see e.g.~\cite{Guth:2007ng,Winitzki:2006rn} for recent reviews).  As a first step, however, we might focus on a single type of bubble, under the approximation that bubble collisions `paint', but do not actually disturb, the interior. The original open FRW foliation allows a calculation of the probability that a given point inside the bubble has various types of collisions to its past.  Morever, in this picture it seems reasonable to equate relative frequencies of observers with relative frequencies of physical volume on a given (homogeneous) equal-time slice.  This is the approach taken in most previous work~\cite{Garriga:2006hw,Aguirre:2007an,Freivogel:2009it,Dahlen:2008rd}.

Here we will review this `core model' and what it indicates about the relative frequencies of (a) the number of bubbles in observers' past lightcones, (b) the observed angular size of those bubbles, and (c) the collision parameter $z_c$ (discussed in Sec.~\ref{sec:thinwallcollisions}). Following this analysis, we briefly describe how including the back-reaction of collisions on the geometry could affect the existence of observers, their likely locations, and what they might see.

\subsection{Probabilities in an undisturbed bubble}\label{sec:coremodel}

Assuming -- as per the thin wall and small bubble approximations -- that the bubbles spread as a null cone emanating from a pointlike nucleation, the total number of collisions expected to be in the past of an observer inside of a bubble is given by
\begin{equation}\label{eq:Ncoll}
N = \lambda V_4,
\end{equation}
where $\lambda$ is the nucleation rate defined in Eq.~\ref{eq:decayrate}, and $V_4$ is the four-volume from which a colliding bubble could be nucleated.\footnote{Of course, in general there might be several types of bubbles that nucleate with different rates from different allowed four-volumes.} An appropriate set of constraints bounding this four volume are that: 
 \begin{enumerate}
 \item  it is outside of the observation bubble -- or else it would represent a decay of the observation bubble's vacuum.
 \item  it is to the past of the observer in question -- so that it is not what we have termed a non-existent bubble.
 \item  it is outside of the past lightcone of the observation bubble's nucleation point -- so that the observation bubble nucleated from the false vacuum rather than another bubble, and 
 \item  it lies to the future of some initial value surface on which the field is assumed to be everywhere in the false vacuum -- so that as discussed below, the false vacuum has not already completely decayed.
 \end{enumerate}

\subsubsection{Reference frames}
\label{sec-refframes}

To set the stage for the problem, consider some region in the false vacuum, from which a single-type of true-vacuum bubble can nucleate. As time progresses, the physical volume in the false vacuum exponentially increases, and different Hubble volumes undergo transitions to other vacua. As discussed in, e.g., ~\cite{Vilenkin:1992uf,Aguirre:2001ks,Aguirre:2003ck} the bubble distribution approaches a steady-state, in the sense that there is a preferred time-slicing in which the statistical distribution of false-vacuum is the same around any given false-vacuum point, and independent of time.  This preferred frame also picks out a `center' to each bubble, given by the worldline that is `at rest' in the preferred frame. Following GGV, we shall assume that, along with the overwhelming majority of bubbles, the observation bubble forms at very `late times', i.e. as part of the steady-state.  This can be implemented by coordinatizing the background false-vacuum with flat spatial slices labeled by time $t$, and assuming that the `initial value' false-vacuum surface is at $t \rightarrow -\infty$, while the observation bubble nucleates at $t=0$.

\begin{figure*}[htb]
\includegraphics[width=14 cm]{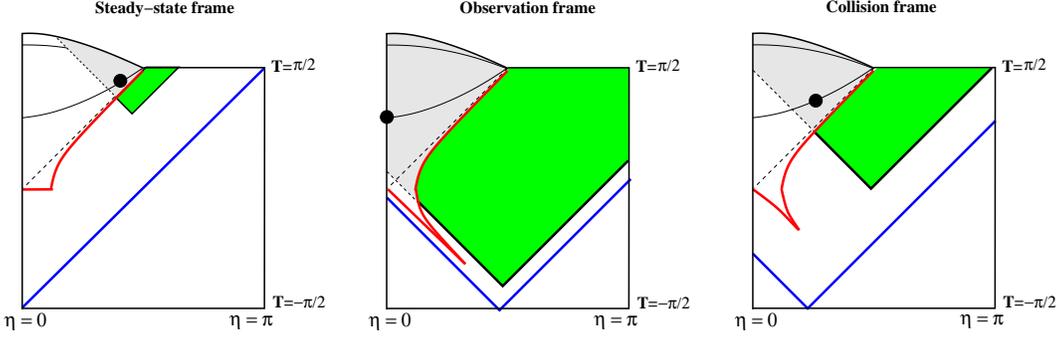}
\caption{The three frames described in Sec.~\ref{sec-refframes} as shown in a `conformal slice' of constant $\theta,\phi$. The observation bubble is on the left, and the colliding bubble on the right. A hypothetical observer is denoted by the dot, and regions inside of the observation bubble to the future of the collision event are shaded grey. The initial value surface (along the $(\theta=0,\phi=0)$ direction separating the nucleation centers) in the background false vacuum is indicated by the solid blue line. In the limit where the interior of the observation bubble remains undisturbed, the transformation between frames is accomplished by boosting in the embedding space, which moves points in the background false vacuum. This brings the colliding bubbles to earlier times, and stretches the observation bubble wall below $T=0$.
  \label{fig-frames}
}
\end{figure*}

The boost invariance of an undisturbed bubble (mentioned in in Sec.~\ref{sec-genericcollision}) is quite useful in that it allows us to define a global transformation that leaves the overall physics unaffected but translates an observer along an equal-time slice within the observation bubble.  As detailed below, this transformation can be described as a global `boost' (of a Minkowski embedding space) that switches between different frames useful for different purposes. Three key frames, illustrated in Fig.~\ref{fig-frames}, are:
\begin{itemize}
\item {\bf Steady-state frame:} This is the frame at rest with respect to the steady state distribution of bubbles in the eternally inflating false vacuum. This is the frame in which the initial value surface is specified and is most useful for connecting bubbles to the background eternally inflating spacetime.
\item {\bf Observation frame:} This is the frame defined by assuming the observer is at rest, and at the center of the observation bubble, with open slicing coordinates $\xi_o=0$. In transforming from the steady-state frame to this one, the initial value surface is distorted as illustrated in Fig.~\ref{fig-frames}.
\item {\bf Collision frame:} In this frame, both the observation bubble and a colliding bubble are nucleated at global slicing time\footnote{To avoid too many coordinate systems in this review, we have made little mention of the coordinates; their properties can be seen in, e.g., AJS or AJT, and the metric for dS in these coordinates is given by Eq.~\ref{eq-globalmetric}.} $T=0$. The initial value surface is distorted from the steady-state frame, and the observer will generally not be at the origin ($\xi_o \neq 0$).  This frame, employed throughout Sec.~\ref{sec-genericcollision}, is most useful for computing the results of a single bubble collision.
\end{itemize} 

To implement the transformation between these frames we embed\footnote{For a disturbed bubble or bubble collision, the spacetime is generally not embeddable in 5D, and there is not a known and simple procedure by which to implement the tranformation, though its character is presumably similar.} the bubble and background dS in Minkowski space with coordinates $X_i, i=0..4$, where $-X_0^2+\vec X^2=H_F^{-2}$ describes the embedded false-vacuum background dS. (See, e.g. GGV, AJ Sec. IIIB, and AJT Sec. IV for details.) In the small-bubble approximation, the observation bubble wall is described by the constant position $X_4=X_{4}^{\rm wall}$. 

We can transform between the various frames by the boost:
\begin{eqnarray}
\label{eq-embeddingboost}
&& X_{0}' = \gamma \left(X_{0} - v X_{1} \right), \\ \nonumber
&& X_{1}' = \gamma \left(X_{1} - v X_{0} \right), \\ \nonumber
&& X_{2,3,4}' = X_{2,3,4}.
\end{eqnarray}
where the steady-state frame is defined by the unprimed coordinates, and different boost parameters $v$ (with  $\gamma=(1-v^2)^{-1/2}$) connect this frame to the collision and observation frames. In the latter case, the observer and $\theta=0$ are both assumed to lie along the $X_1$ direction from the origin, and the boost parameters $\gamma_o = \cosh \xi_{o}$ and $v_o = \tanh \xi_{o}$ translate an observer at $\xi_o$ to the origin of the observation bubble, while shifting the position of the colliding bubble. In the former case we assume that the colliding bubble's nucleation center is along the $X_1$ direction, then find the boost such that the nucleation is at global slicing time zero. (See AJ Sec. IIIB for the boost parameters.)
The observer in this frame is generally located at an open slicing position  $\xi_o ' \neq 0, \theta_o \neq 0$.

\subsubsection{Expected number of collisions}
\label{sec-expnum}

\begin{figure*}
\includegraphics[width=15cm]{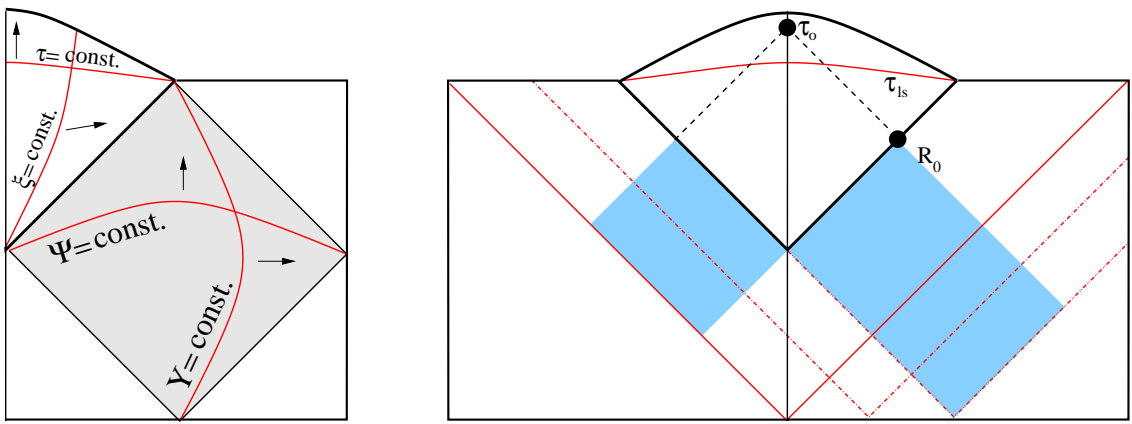}
\caption{The coordinates defined in Eq.~\ref{eq:outsidehypercoords} cover the shaded region of the false vacuum de Sitter space in the left panel. Lines of constant $\Psi$ and $\Upsilon$ are indicated in the region outside the bubble, and lines of constant $\xi$ and $\tau$ inside of the bubble. In the right cell, we show the general setup for the core model. The initial value surface for observers at increasing $\xi_o$ is boosted from the solid line in the exterior false vacuum to the dashed lines, as in Fig.~\ref{fig-frames}. Tracing the past light cone back from the observer at $\tau_o$, it intersects the surface of last scattering at $\tau_{\rm ls}$, then the bubble wall (which at this time has radius $R_0$), and finally passes into the exterior false vacuum. The portion of the false vacuum that is available for the nucleation of colliding bubbles in the past light cone of the observer is shaded.  
\label{fig-volumes}}
\end{figure*}

The general setup is illustrated in Fig.~\ref{fig-volumes}, which shows a conformal slice ($\theta = 0, \phi=0$) through the single bubble spacetime in the observation frame. The initial value surface for $\xi_o = 0$ is located at flat slicing $t=-\infty$, and we show the initial value surface for increasing $\xi_o$ (i.e. after the boost into the observation frame, as specified below). Any bubbles nucleating from the shaded region between the bubble wall and the initial value surface produce a potentially observable collision.

Within this picture, convenient coordinates on the false vacuum de Sitter space are~\cite{Freivogel:2009it}:
\begin{equation}\label{eq:outsidehypercoords}
ds^2 = \frac{1}{H_F^2 \cosh^2 \Upsilon} \left( -d \Psi^2 + d\Upsilon^2 + \cosh^2 \Psi d\Omega_2^2 \right).
\end{equation}
These are induced by the embedding:
\begin{eqnarray}
&& X_{0} = H_{F}^{-1} \frac{\sinh \Psi}{\cosh \Upsilon};\ \ \  X_{i} = H_{F}^{-1} \frac{\cosh \Psi}{\cosh \Upsilon}  \omega_i , \\ \nonumber
&& X_{4} = - H_{F}^{-1} \tanh \Upsilon,
\end{eqnarray}
where $- \infty < \Upsilon < \infty$ and $-\infty < \Psi < \infty$. These coordinates are illustrated in Fig.~\ref{fig-volumes}; note that the the bubble wall is located at $\Upsilon \rightarrow -\infty$, that $t\rightarrow -\infty$ in the flat slicing coordinates corresponds to $\Psi =\Upsilon$, that the past lightcone of an observer at the origin is given by $\Psi = -\Upsilon+(const.)$, and that boosts preserve $\Upsilon$ while changing $\Psi$.

Now, orienting the coordinates so that our observer is at $\theta=0, \xi'=\xi_o$ in the (primed) steady-state frame, we can transform to the observation frame so that the observer is at $\theta=0, \xi=0$.  The observer's past lightcone can then be found by matching its radius $R_0$ (given for a realistic bubble cosmology in Eq.~\ref{Rosimp}) across the bubble wall to find
\begin{equation}\label{eq:PLCinFV}
\Psi = - \Upsilon + \log \left[ H_F R_0 \right].
\end{equation}

The initial value surface, given in the steady state frame by $\Psi' = \Upsilon'$, is transformed by the boost, which sets $\Upsilon = \Upsilon'$ and transforms $\Psi$ to give:
\begin{equation}\label{eq:boostedivsurface}
\sinh \Psi = \gamma \left( \sinh \Upsilon - v \cosh \Upsilon \cos \theta \right)
\end{equation}
where $\gamma = \cosh \xi_o$ and $v = \tanh \xi_o$.

\begin{figure*}
\includegraphics[width=18cm]{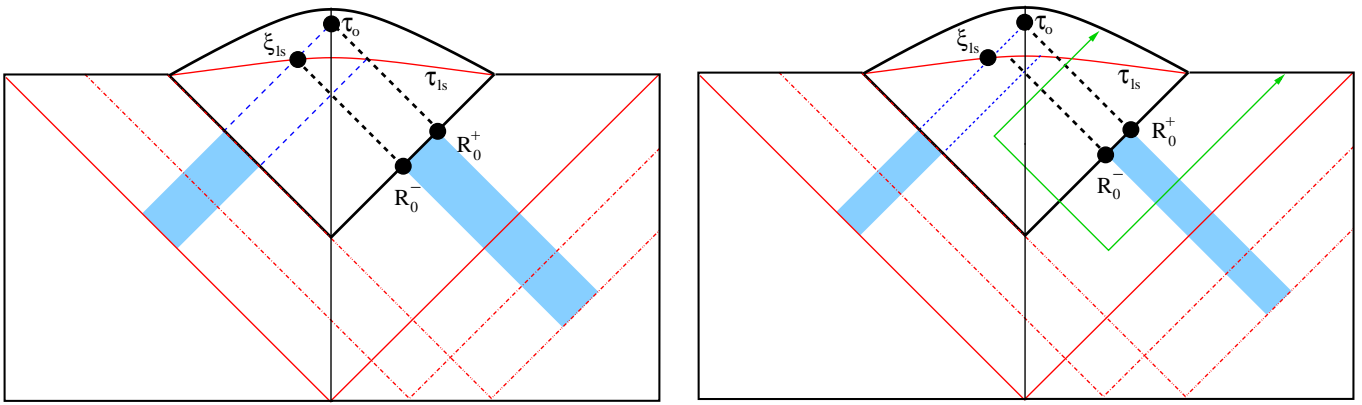}
\caption{In the left panel, we shade the four-volume from which colliding bubbles can produce effects that do not encompass the entire CMB sky. This is found by tracing back null rays from the visible portion of the surface of last scattering into the false vacuum (see Appendix~\ref{sec-nucpastcalc} for the quantitative details). Note that the four-volume to the future of the initial value surface near past-infinity is now excluded, regulating the divergent number of expected collisions at large $\xi_o$. In the right panel, the four-volume satisfying the criteria outlined in Ref.~\cite{Freivogel:2009it} is shaded. Here, any collisions that could produce a post-collision domain wall intersecting the origin are excluded. One such collision is depicted by the green line with arrows. Again, the potentially allowed four-volume near past infinity is regulated. 
\label{fig-volumes2}}
\end{figure*}

With this setup, the number of nucleations to the past of an observer at position $\xi_o$ can be calculated, as detailed in Appendix~\ref{sec-nucpastcalc}. The basic results can be described qualitatively using Fig.~\ref{fig-volumes}, and quantitatively using the results of that Appendix, as follows:
\begin{itemize}
\item At $\xi_o = 0$, the shaded area in the right panel of Fig.~\ref{fig-volumes} can become large if $\tau_o$ becomes large compared to $H_F^{-1}$, so that the shaded areas start to include the region near future infinity of the false vacuum. Because most of these collisions enter the past light cone of an observer at large $\tau_o$, these were denoted `late-time' collisions by AJS. Quantitatively, if there are $N_e \agt 1$ e-folds of inflation, the expected number of collisions is
\begin{equation}\label{eq:numbernoboost}
N \simeq \frac{4 \pi \lambda}{3H_F^{4}} \left( \frac{H_F^2}{H_I^2}\right),
\end{equation}
with logarithmic corrections (see Eq.~\ref{eq-nxi0}).

\item At increasing $\xi_o$, there is more shaded area to the right of the observer in Fig.~\ref{fig-volumes} than to the left, indicating that the observer records more collisions coming from the $\theta=0$ direction than its opposite (or in fact any other direction). This anisotropy represents a `persistence of memory,' first described by Garriga, Guth, and Vilenkin (GGV)~\cite{Garriga:2006hw}, whereby the effects of the initial value surface on the statistics of collisions are not erased, even when the initial value surface is infinitely far to the past of an observer.  Moreover,  the distortion of the initial value surface leads (in some directions) to an increase in the four-volume near past infinity of the false vacuum, and consequently an increase in the expected number of collisions. These collisions enter the past light cone of an observer at very small $\tau_o$, and were therefore denoted as `early-time' collisions by AJS. Quantitatively, for $\xi_o \rightarrow \infty$, the total number is given by
\begin{equation}
\label{eq-simpleprob}
N \simeq \frac{4 \pi \lambda }{3H_F^4} \left(\frac{H_F^2}{H_I^2} \right) \xi_o.
\end{equation}

\item The total counts given in Eqs.~\ref{eq:numbernoboost} and~\ref{eq-simpleprob} include many collisions that encompass the entire CMB sky, where the null shell of collision debris does not intersect the visible portion of the surface of last scattering. In fact, the vast majority of the divergent number of expected collisions in Eq.~\ref{eq-simpleprob} are of this type. If we only count collisions that do not encompass the entire CMB sky, the relevant four-volume is shown in the left panel of Fig.~\ref{fig-volumes2}. Note that the region near past infinity at large boost is no longer included, which implies that the position of the observer inside of the bubble is largely irrelevant and the divergent number of early-time bubbles is not counted. This disappearance of the dependence on the initial value surface was termed by FKNS as the `disappearance of the persistence of memory.' In Appendix~\ref{sec-nucpastcalc}, we calculate the expected number of collisions to be (in the small $\xi_{\rm ls}$ limit)
\begin{equation}\label{eq:numberintersect}
N \simeq \frac{16 \pi \lambda}{3H_F^{4}} \left( \frac{H_F^2}{H_I^2}\right) \sqrt{\Omega_c}
\end{equation}

\item An alternative criteria, employed by FKNS, is to forbid any bubbles for which the post-collision domain wall crosses the (putative) worldline of the observer. Assuming that the domain wall travels into the bubble for an inflationary Hubble time, and then accelerates outward, the relevant four-volume is sketched in the right panel of Fig.~\ref{fig-volumes2}. The number of collisions satisfying this criteria is given roughly by Eq.~\ref{eq:numberintersect} up to factors of order unity (again in the small $\xi_{\rm ls}$ limit).

\end{itemize}

\subsubsection{Angular scale distribution}

While the number of nucleations to the past is important, equally vital is the area on the observer's sky that might be affected by the collision: solid angles too close to $0$ are clearly bad news, and bubbles covering $4\pi$ might be interesting, but might also be invisible if the whole sky is effectively seeing just a tiny (and hence potentially homogeneous) part of the full collision spacetime. In Sec.~\ref{sec:angularscale}, we provided a formula (Eq.~\ref{eq:costhetadS}) relating the nucleation site of a colliding bubble to the observed angular scale on a surface of constant $\tau$. Using the volume element Eq.~\ref{eq:dV4} and the available nucleation four-volume described in the previous section, we can calculate the statistical distribution of observed angular scales on, for example, the CMB sky. The detailed calculation is given in Appendix~\ref{sec-angscalecalc}; we summarize key results here.

\begin{figure}
\includegraphics[width=8cm]{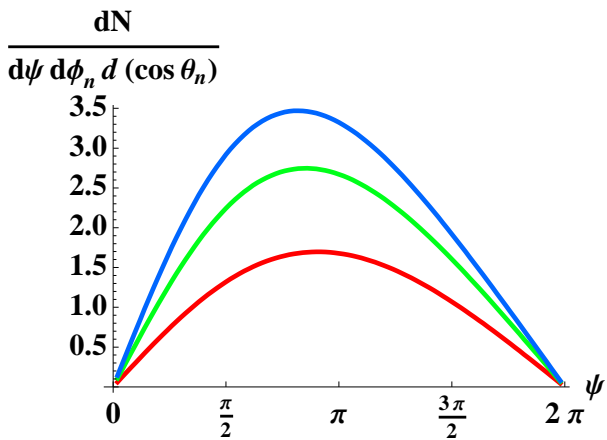}
\caption{The distribution function Eq.~\ref{dndpsi} normalized by $\lambda H_{F}^{-4}$. We chose the parameters $N_e = 70$, $\frac{H_{F}}{H_{I}} = 10$, $\theta_{o}=0$, and $\xi_o \rightarrow \infty$ with $\xi_{\rm ls} =.05$ shown in red (bottom), $\xi_{\rm ls} =.08$ in green (middle), and $\xi_{\rm ls} =0.1$ in blue (top). The maximum scales like $\left( \frac{H_F}{H_I} \right)^2 \xi_{\rm ls} $, and as $\xi_{\rm ls} \rightarrow 0$, the shape approaches $\sin \left( \frac{\psi}{2} \right)$.
\label{fig-rhdistribution}}
\end{figure}

\begin{figure*}
\includegraphics[width=8cm]{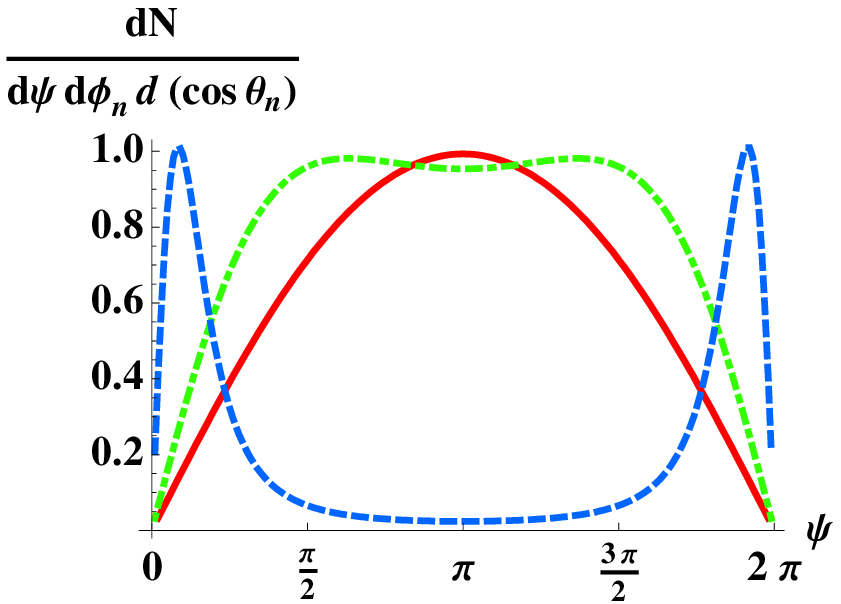}
\includegraphics[width=8cm]{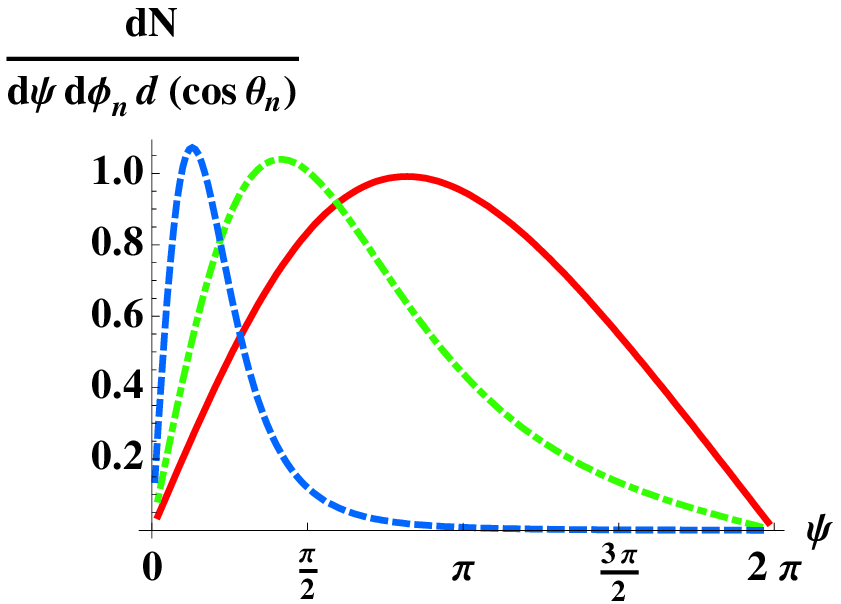}
\caption{The distribution function Eq.~\ref{dndpsi} for $\frac{H_{F}}{H_{I}} = 1$ (left) and $\frac{H_{F}}{H_{I}} = 10$ (right) with increasingly large $\xi_{\rm ls} =$0.1 (red, solid), 0.5 (green, dot-dashed), and 2 (blue, dashed).  For all curves, $N_e = 70$, $\theta_{o}=0$, and $\xi_o \rightarrow \infty$. We have scaled out the increasing height of the peak that comes with increasing $\xi_{\rm view}$. The distribution becomes bimodal for small $\frac{H_{F}}{H_{I}} $, and obtains a large peak near $\psi \sim 0$ for large $\frac{H_{F}}{H_{I}} $.
\label{fig-distributionlargexi}}
\end{figure*}

\begin{itemize}

\item In general, the angular distribution depends upon the nucleation direction $\theta_n, \phi_n$, the comoving radius $\xi_{\rm ls} \sim \xi_{\rm reh}$ out to which the observer can see on the surface of last scattering, the false and true vacuum Hubble constants $H_F$ and $H_I$, and the number $N_e$ of inflationary e-foldings inside the observation bubble.

\item We can gain some intuition by comparing the map of angular scales in Fig.~\ref{fig-anglesmap} to Figs.~\ref{fig-volumes} and~\ref{fig-volumes2} depicting the relevant nucleation regions. The distribution will have strong support at some angular scale if the corresponding map region overlaps with an allowed nucleation region of large four-volume. In the limit where $\xi_{\rm ls}$ is very small, the map is quite compressed, and each range of angular scales receives a nearly equal contribution. We therefore expect the angular scale distribution to be rather flat in this case. Considering larger $\xi_{\rm ls}$, regions near to future infinity (for all $\xi_o$) and past infinity (in the case where the observer is at large $\xi_o$) are on the map's boundaries near $\psi \sim 0$ and, depending on the viewing angle, $\psi \sim 2 \pi$. For large $\xi_{\rm ls}$, there should then be a bimodal distribution strongly peaked around small and large angular scales~\footnote{Recall from the previous subsection that at small $\xi_{\rm ls}$, the majority of collisions to the past of an observer cover the whole sky. As $\xi_{\rm ls}$ increases, more of these bubbles are included in the count, leading to the peak around large angular scales. }. The peak about $\psi \sim 0$ will be isotropic, and the peak about $\psi \sim 2 \pi$ will be anisotropic and maximized about $\theta_o = 0$. If the ratio $H_F / H_I$ is large, then the peak near $\psi \sim 0$ will be larger than the peak near $\psi \sim 2 \pi$ because more volume is included near future infinity than past infinity, leading to a bias for small scale collisions.

\item As discussed in Sec.~\ref{sec-onebubreview}, for a realistic cosmology inside of the bubble $\xi_{\rm ls}$ is rather small. In Fig.~\ref{fig-rhdistribution}, we show the numerically computed distribution function normalized by $\lambda H_{F}^{-4} $ for the case where $N_e = 70$, $\xi_{\rm ls} = \{ .05, .08, .1\}$, $\frac{H_{F}}{H_{I}} = 10$, $\theta_{o}=0$, and $\xi_o \rightarrow \infty$. Choosing different observation angles has little effect on the shape of the distribution function, meaning that the observed bubbles will be distributed fairly isotropically. In accord with our expectations, the distribution is rather flat, and approaches
\begin{equation}
\label{eq-angdist}
\frac{dN}{d\psi d(\cos \theta_n) d\phi_n} \sim \lambda H_F^{-4} \left( \frac{H_F}{H_I} \right)^2 \xi_{\rm ls} \sin \left( \frac{\psi}{2} \right).
\end{equation}
in the limit where $\xi_{\rm ls} \rightarrow 0$. It can be seen from the numerical distribution that this is a good assumption for $\xi_{\rm ls} \alt 0.05$. This distribution was first derived by FKNS using different methods. 

\item As $\xi_{\rm ls}$ is increased, the distribution changes as shown in Fig.~\ref{fig-distributionlargexi}. As expected from the discussion above, when $\frac{H_{F}}{H_{I}}$ is ${\cal{O}}(1)$, the distribution becomes bimodal in the limit of large $\xi_o$ due to the inclusion of an equal amount of four-volume near past and future infinity (left panel of Fig.~\ref{fig-distributionlargexi}). For large $\frac{H_{F}}{H_{I}}$, there is more four-volume near future infinity, weighting the distribution towards small angular scales. The peaking of the distribution near $\psi \sim 0$ and possibly $\psi \sim 2 \pi$ is directly related to the ability to view approximately one curvature radius on some early-$\tau$ surface. A similar bimodal distribution function was found in AJS for the angular scale of collisions on the bubble wall. Those results can be reproduced from the analysis in this section by taking the appropriate limit where $\tau_{\rm view} \rightarrow 0$. 
\end{itemize}

\subsubsection{Statistical distribution of $z_c$}
\label{sec-zcdist}

In Sec.~\ref{sec:thinwallcollisions} we defined $z_c$, the value of $z$ (in the coordinates of Eq.~\ref{metric}, in the collision frame) at which the bubbles first collide. Under the assumptions and approximations we employ, this is the sole parameter -- beyond those determined by the inflaton potential and other microphysics -- that  determines the resulting collision spacetime. The quantity $z_c$ is directly related to the (boost) invariant separation of the nucleation centers of the bubbles, which (in the small-bubble approximation) is simply given by the value of $\Upsilon$ at which the colliding bubble nucleates. Therefore, in just the same manner as the derivation of the angular scale distribution, we can find the distribution in $z_c$ by integrating the volume element. 

\begin{figure}
\includegraphics[width=8cm]{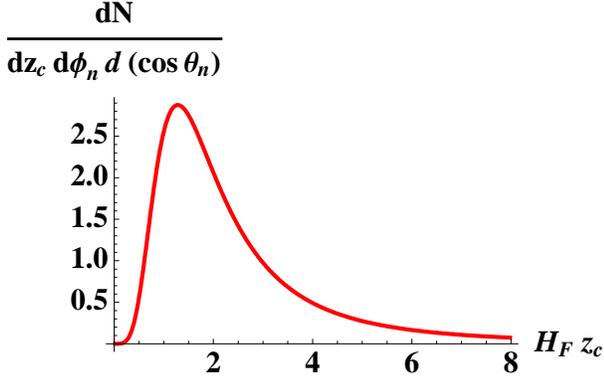}
\caption{The expected number of collisions with various $z_c$, the value of $z$ at the collision in the collision frame. Note that the distribution peaks near $z_c \sim H_F^{-1}$. The intra-bubble cosmology is specified by $N_e = 70$, $\frac{H_{F}}{H_{I}} = 10$, and $\xi_{\rm ls} =.05$. The observation angle and position of the observer are set to $\theta_{o}=0$ and $\xi_o \rightarrow \infty$.
\label{fig-zcdist}}
\end{figure}

This analysis was performed in AJ, and we refer the reader there for further details. Changing variables to $z_c$, and generalizing the treatment in AJ to include only collisions intersecting the observable portion of the surface of last scattering (this makes little difference for the shape of the distribution), we show the numerically calculated distribution function for $N = 70$, $\frac{H_{F}}{H_{I}} = 10$, $\theta_{o}=0$, $\xi_o \rightarrow \infty$, and $\xi_{\rm ls} =0.05$ in Fig.~\ref{fig-zcdist}. The distribution peaks around $z_c \sim H_F^{-1}$, and falls off quickly at large $z_c$: a quantitative examination of the distribution (after straightforward algabraic manipulations of the expressions in AJ) shows that the fall-off at large $z_c$ goes like 
\begin{equation}
\frac{dN}{d z_c d\phi_n d\theta_n} \simeq \lambda H_F^{-4} \left( \frac{H_F^2}{H_I^2} \right) \frac{8 \xi_{\rm reh}}{H_F^3 z_c^3}
\end{equation}
As a fraction of the height of the maximum, this is roughly $H_F^{3 }z_c^3$. 

Thus for typical collisions, values $z_c \sim H_F^{-1}$ should be assumed when analyzing the effects of collisions.  For example, assessing the strength of the observable effects of collision debris presented in Sec.~\ref{sec:observingdebris}, only exceedingly rare collisions would be at very large $z_c$ where the strength of effects associated with the collision debris can become large.

\subsection{Volume fractions}

\begin{figure}
\includegraphics[width=7.5cm]{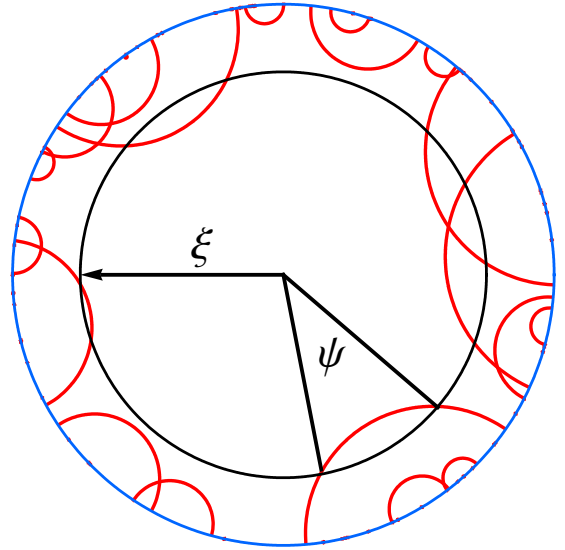}
\caption{The Poincare disc representation of a constant $\tau$ surface inside of a bubble undergoing collisions. Regions enclosed by the red `tongues' represent portions of the surface of constant $\tau$ that are to the future of a collision event. To generate this picture, the sizes and locations of the tongues were chosen randomly from an angular scale distribution function such as the one outlined in the previous section. Bubbles within bubbles and regions with overlapping collisions were not subtracted here, although they might be in a more accurate treatment. 
\label{fig-consttau}
}
\end{figure}

The probability distributions calculated in the previous section pertain to an observer at a given position inside the observation bubble, and are essentially determined by the probability for bubble nucleation and the details of embedding the bubble inside of the false vacuum background. We now take a more `global' viewpoint and calculate the volume fraction in the future of various collision events on a surface of constant $\tau$. This could be taken as a measure, in which the relative frequencies of different possible observations are given by these volume fractions (though of course other measures are possible). In Fig.~\ref{fig-consttau}, we show one such surface of constant $\tau$ in the `poincare disc' representation of an open FLRW universe (see, e.g., AJS for metric and embedding of this representation). The boundary of the disc corresponds to $\xi \rightarrow \infty$, and red discs (colloquially denoted `tongues' in some of the literature) are drawn to signify the regions affected by collisions.

A reasonable way to compute overall volume fractions on a surface of constant $\tau$ is to compute them on a ball of constant $\xi$, then take the limit $\xi \rightarrow \infty$, since the overall volume is dominated by large $\xi$.  In terms of the volume fraction affected or unaffected by collisions, one can do this by directly calculating the solid angle on the constant-$\xi$ sphere~\cite{Garriga:2006hw,Dahlen:2008rd} free of collisions. Equivalently, we can use the fact that the probability for a point on the sphere {\em not} be to the future of a collision is given by~\cite{Guth:1981uk}
\begin{equation}
P = e^{- \lambda V_4}
\end{equation}
where $V_4$ is the four-volume to the past of a given point, which at large-$\xi$ is Eq.~\ref{eq:case1v4largexi}. 

The total volume fraction unaffected by collisions is then
\begin{equation}\label{eq:volumefractions}
\frac{V_{un}}{V_{total}} = \frac{\int d\xi \sinh^2 \xi e^{- \frac{4 \pi}{3 H_F^2 H_I^2} \lambda \xi } }{\int d\xi \sinh^2 \xi } \sim e^{- \frac{4 \pi}{3 H_F^2 H_I^2} \lambda \xi_{\rm max}}.
\end{equation}
Therefore, even though the volume unaffected by collisions grows without bound as $\xi_{\rm max} \rightarrow \infty$, the fraction of unaffected volume goes to zero.

Nevertheless, the size of a typical `pristine' region inside of the bubble can be quite large. We can estimate the distance in $\xi$ out to which one must go from a pristine point to encounter a collision by calculating the four-volume to the past of a disc of radius $\xi$ on a surface of constant $\tau$, excluding the four-volume to the past of the central point. This calculation was performed by GGV in the case where $H_F = H_I$. If this is not the case, the four-volume straightforwardly generalizes to
\begin{eqnarray}
V_4 &=& \frac{4 \pi}{H_F^2 H_I^2} \xi, \ \ \ (\xi \ll 1), \\
V_4 &=& \frac{4 \pi}{3 H_F^2 H_I^2} e^{2 \xi}, \ \ \ (\xi \gg 1),
\end{eqnarray}
in the limit where $N_e > 1$. For very small nucleation rate, the expected distance to the first collision is then
\begin{equation}
\Delta \xi \sim \log \left( \frac{H_F^2 H_I^2}{\lambda} \right).
\end{equation}
This can be many curvature radii, allowing for regions many times the size of our observable universe to exist between collisions. However, for $\lambda H_F^{-4} \sim H_I^2 / H_F^2$, we would expect regions the size of our observable universe to contain at least one collision, in agreement with the previous calculations of the expected number of collisions.

A more complex analysis might split the region to the future of the collision into different categories, for example into those regions inside and outside of a post-collision domain wall. Collisions with bubbles of the same vacua requires such considerations, since the bubble interiors merge, and one must define which bubble a given observer is in~\cite{Dahlen:2008rd}. Following Dahlen~\cite{Dahlen:2008rd}, in a collision between two identical bubbles, we can define the region still inside the original bubble by going to the collision frame, and including only the volume contained within $ x< x_c$, where $x_c$ defines the line of symmetry across which the bubbles intersect. Because most of the volume affected by the collision is at large $\xi$, where this procedure excises a large part of the collision region, it is interesting to calculate the fraction of volume in the future of a collision (but still within the original bubble) to the unaffected volume. The result of this computation is that the volume fraction (and by the volume weighting measure, the probability to be) to the future of a collision is given for small nucleation rates by
\begin{equation}\label{eq-dahlenvolfrac}
f \simeq \frac{4 \pi \lambda}{3 H_F^2 H_I^2}.
\end{equation}
It would be instructive to repeat this computation in the case where the bubbles are not identical, though the result will depend sensitively on the time at which the fraction is evaluated, since the domain wall is accelerating and thus divides the collision region differently on different constant $\tau$ surfaces.

\subsection{Modifications from back-reaction by bubble collisions}

In previous sections, we have worked in the approximation that collision effects simply `paint' the constant-$\tau$ surfaces of an undisturbed bubble. What might a more realistic treatment of the bubble interior, including the effects of collisions, mean for the likelihood that we might observe a collision? Giving up the symmetry of a homogeneous bubble interior immediately opens the door to a host of complications as well as difficult issues of principle (e.g. the measure problem) for which there are no clear answers. Our discussion is therefore much less concrete than in previous sections, and primarily serves to illustrate some of the possibilities.

One of the key results of Sec.~\ref{sec-genericcollision} was that when the post-collision domain wall accelerates away from the observation bubble, and the collision does not disrupt inflation, then it is possible to recover infinite spacelike reheating surfaces with approximate homogeneity and isotropy to the future of the collision. One possible way to represent such a collision spacetime is in terms of surfaces of constant inflaton field or density. In the limit where such surfaces are everywhere spacelike, the collision effect would appear as a minor distortion of the `painted Poincare disc' representation above. If the surfaces go timelike over some interval, such a representation may still be possible by including a narrow `junction region' of inhomogeneous field (or density) joining the two uniform regions. We can also include the effects of all other collisions -- where there are no observers to the deep interior -- in this picture as a black `bite' taken out of the disk. These would include (a) collisions in which the domain wall accelerates inwards, (b) collisions in which there is very little inflation, or (c) collision with an identical bubble, so that much of the volume is attributed to the other bubble, as above.

\begin{figure*}
\includegraphics[width=10cm]{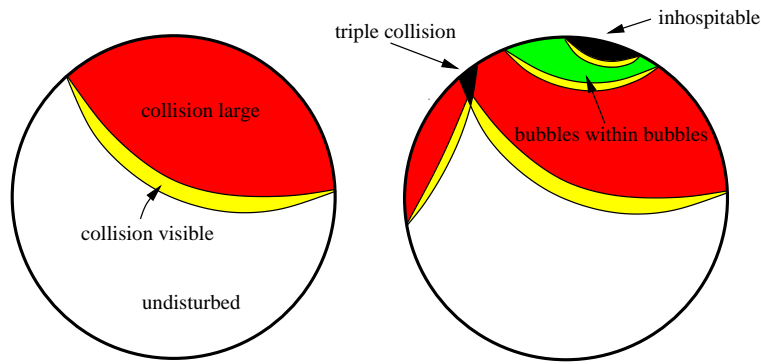}
\caption{In the case where homogeneity and isotropy are restored to the future of the post-collision domain wall, we can again think of the collisions as dividing the original Poincare disc representation of a constant density hypersurface into different regions. As shown in the left panel, there will be three types of regions to consider: the region undisturbed by the bubble collision, the banana-shaped boundary region in the neighborhood of the collision where the effects of the interface might be visible, and the deep interior region to the future of the collision where the interface is not visible. In the red region, subsequent collisions can occur with bubbles nucleated within bubbles, possibly producing a different boundary and deep interior (green). This nesting of collisions within collisions will occur until, for example, the cumulative effect of many collisions ruins the prospects for inflation (black region). Further effects, such as multiple collisions, could also be important for the structure of various regions. 
\label{fig-disc2}
}
\end{figure*}

Such a representation of (for example) the reheating surface is shown in Fig.~\ref{fig-disc2}. For a single collision, there are two regions of interest: the boundary of the collision region (yellow bananas) and the `deep interior' of the collision region (red lozenges). We might define the deep interior as the region homogenous and isotropic enough to support foreign-born observers (in the language of Sec.~\ref{sec:obssignatures}), with the banana representing the remaining region affected by the collision.

Deep interior regions are to the future of a post-collision domain wall, and therefore have the colliding bubble, rather than the  `original' false vacuum to their past. These regions are largely shielded from subsequent collisions with bubbles nucleated out of the (original) false vacuum, but still vulnerable to collisions with bubbles nucleated within the {\em colliding} bubble. Then, recapitulating the exact large-$\xi_o$ argument pertaining to the original observation bubble, if we consider regions far enough into the `deep interior', they are sure to experience such a collision. Again, if the post-collision environment of such collisions allows for inflation and late-time infinite constant field hypersurfaces, there will be a boundary region and an deep interior (green lozenge in Fig.~\ref{fig-disc2}); otherwise there will be a black `bite' out of the lozenge.

With this in mind, we can think of the overall structure of the original observation bubble and its distribution of observers, as follows.  Beginning at the center, let us move towards the edge of the disk and consider the regions we cross over that can support observers, keeping in mind the rough intuition that regions nearer the edge of the disk overwhelmingly dominate the disks's volume. We start with observers in the observation bubble to the future of the false vacuum.  We eventually enter a banana, then a collision region's deep interior region to the future of the collision bubble.  Continuing, we enter the border then deep interior to the future of a child of the collision bubble, etc.

This nesting of collision regions containing observers continues until either (a) the last collision is stable to subsequent bubble collisions, or (b) the last collision is a black-bite region with no observers. In the first case, there are no more transitions to occur at greater radius, and since the volume is dominated by the largest radii, we might expect regions like this to contribute overwhelmingly to the overall volume of regions that support observers.

Weighting by either physical or comoving volume, and assuming that something akin to the open FLRW foliation holds in the pasted-together collision spacetime so that the volume is dominated by regions near the disk's edge, we can use the results of the previous section to highlight some of the possiblitites:
\begin{itemize}
\item The volume fraction in the boundary bananas, where the effects of collisions are observable, is small unless the nucleation rate $\lambda_B H_B^{-4}$ out of some vacuum is large relative to $(H_B2/H2_I)$, where $H_B$ corresponds a vacuum able to nucleate bubbles that reach the region in question.  This follows from Dahlen's result Eq.~\ref{eq-dahlenvolfrac} for identical bubble collisions described above. By similar reasoning, among the different types of boundary regions, we would expect most of the boundary volume to be to the future of the colliding bubble with the highest value of $\lambda_B/H_B^2H_I2$.
\item Regions that collide with a stable bubble are rewarded, and if they allow observers, may dominate the ensemble of observers.
\item Second in importance to these are the deep interiors of the last regions (moving towards the disk's edge) to support observers.
\item It is likely that the  `painted disk volume' is dominated by the black inhospitable regions.  It is, however unclear exactly to what degree this is a problem: these regions are sufficiently different in their spacetime structure at later times (when observers might arise) that it is hard to compare them to the inflationary regions.
\end{itemize}

The last point once again brings in the nefarious measure problem, about which we will make two last points here.

First, the effects of collisions could be important for measures over the entire eternally inflating multiverse\footnote{Some attempts to understand the measure problem stem more directly from the statistical properties of collisions, for examples of how this may be relevant see~\cite{Freivogel:2009rf,Bousso:2008as,Freivogel:2006xu}. }, which necessarily must compare observers in different bubbles. One possibility, suggested by CKL and further developed in AJT, is that there may be a preference for bubbles which have a lower Hubble scale, since they are more likely to repel post-collision domain walls. One way of making this precise [AJT] is to define a cutoff scheme that compares the reheating volume in different bubbles. Consider bubble A and bubble B, which are nucleated out of the same false vacuum and can collide. The inflationary Hubble scale in bubble A is lower than in bubble B, and therefore when these two bubbles collide, the post-collision domain wall (in the thin wall limit with an appropriate tension -- see Eq.~\ref{eq:domainwalloutcond}) moves into B. If inflation occurs to the future of the collision inside of bubble A, then very roughly, for every portion of the reheating surface removed from B, a portion is gained inside of A. In the painting approximation, the fraction of volume on the reheating surface of A to B measured out to some comoving distance $\xi$ in each bubble goes like~\ref{eq:volumefractions}, which approaches infinity as you send $\xi \rightarrow \infty$. Therefore, including the effects of collisions may give strong weight to bubbles with a low scale of inflation.

Second, it is also quite possible that the overall measure may strongly impact how we count, and thus that we expect to be seen by, observers in bubbles.  For example, several extant measure proposals such as the `scale factor cutoff' measure~\cite{DeSimone:2008bq} would ascribe most weight to observers at the center of the observation bubble~\cite{Bousso:2008hz} (at rest with respect to the steady-state frame defined by the initial value surface).  In this case, the edge of the Poincare disk representation may be largely irrelevant, and the above discussion could be considerably altered.

\section{Overall summary and discussion}

Eternal inflation and its accompanying bubble or pocket 
universes are in some sense a theoretical byproduct of invoking inflation to explain the boundary conditions for the standard big bang cosmology. This scenario also arises as a side effect of many theories with extra dimensions, such as string theory, which generically give rise to many vacuum solutions. Thus in taking seriously many theories going beyond the standard model of particle physics and cosmology, we must also take seriously the prospect of living in a multiverse, in which thus-far observed physical properties of our world pertain only locally. Given the profound reconceptualization of the Universe that this picture entails, and the very thorny physical and even philosophical issues the idea raises, it is important to determine how, even in principle, we might directly probe such multiverse scenarios experimentally or observationally. In this review, we have explored one such possibility: the observation of relics from the collision between bubble universes. While requiring no small amount of good fortune, the detection of such a signal would truly be an epochal discovery.

The research we have reviewed encompasses many possibilities for the spacetime region affected by a bubble collision, ranging from essentially empty space inconsistent even with observers' existence, to regions indistinguishable from standard inflationary cosmology.  By far the most interesting scenarios are those in between, in which observers might see a cosmology that is similar to, but detectably different from, that described by the current standard cosmological model. If those observers are in some sense generic (relative to observers who see no effect of the collision), then we have a relatively direct way of testing any cosmological model giving rise to such collisions, as well as the exciting possibility of directly detecting evidence of a multiverse.

How fortunate must we be in order to observe a collision?  The arguments and calculations to date indicate that the following must be true:
\begin{enumerate}
\item The inflaton potential must admit bubbles and their collisions, and at least one type of bubble with a cosmology in accord with our observations.
\item {At least one type of bubble collision must allow nearly-normal cosmological evolution to its future; probably the prime requirement for this is
\begin{enumerate}
\item collision with a bubbles of identical vacuum, or with those that forms a domain wall accelerating away from the observer's bubble, in which \item there are sufficiently many e-folds of inflation in the collision region.
\end{enumerate}}
\item There must not be {\em so} many e-folds of inflation that the effects of the collision are stretched into unobservability. Because the effects of a collision can be considered a `pre-inflationary relic' in terms of inflation inside the bubble, roughly speaking there must not be many more e-folds of inflation within the bubble than  necessary to explain, e.g., the small observed cosmic curvature.
\item Several lines of argument suggest that while essentially all locations within a given bubble have collisions to their past, in order for us to {\em detect} collisions, there must be regions to our past with a `false' vacuum that can decay at a rate $\lambda H_F^{-4} \agt (H_I / H_F)^2$ where $H_I$ and $H_F$ are the Hubble parameters during inflation and in the parent vacuum. Roughly speaking, this criterion stems from the fact that given our small observed curvature, we can see only about one inflationary Hubble volume, to the past of which is are roughly $(H_I/H_F)^2$ Hubble four-volumes from which bubbles can nucleate.
\end{enumerate}

Given a specific model of the inflaton potential, it would be possible to check whether these criteria are at least roughly met.  While we currently do not have strong grounds either observationally or theoretically to single out a specific inflaton model, we can try to assess what sort of features might be `reasonable' or not. For example, it appears somewhat awkward to contrive tunneling potentials that are both amenable to semi-classical treatment and also give a tunneling rate that is not very exponentially suppressed (making criterion 4 a challenge).  However, such arguments are, by their nature, never really definitive. 

Another possibility is look for {\em observational} clues that might winnow the possibilities in ways either favorable or not to observable bubbles. For example, the detection of negative curvature would give some indication that we reside in a bubble universe (criterion 1), and that there were a minimal number of e-folds during inflation (criterion 2).  Contrariwise, the observation of positive curvature would rule out the possibility that we inhabit a relatively uniform bubble universe. As another example, if we observe primordial gravitational waves in future CMB polarization experiments, this probably indicates that the inflaton potential was of the large-field type, and therefore that inflation can endure to the future of a collision.

Absence of compelling evidence for just the right sort of inflationary dynamics should not, however, stop us from assessing what a bubble collision would look like observationally, and for looking for it.  Analysis to date suggests  that evidence could come from a number of features. Some examples might include the presence of one or many azimuthally symmetric features in the CMB, perhaps with a hard edge, correlated with other signals like bulk flows or a perturbation power anisotropy in the same direction

Even if collisions are never seen, the possibility of their detection at least provides a way in which particular models of the multiverse can be ruled out. This would be the case if signatures (such as those described above) were not detected in a model where we expect to see collisions with dramatic effects. For example, this could include models that predict fast nucleation rates with few e-folds of inflation to the collision's future.

The ideas presented in this review consider only the simplest scenarios and possible observational effects. There is much room for future work exploring all aspects of the problem, including the structure of the post-collision spacetime, the types and strength of observational effects, the likelihood that we might see a collision in a given model, and the types of plausible landscape scenarios that might give rise to eternal inflation. With some luck, the discovery of `other universes', a concept seemingly out of science fiction, may be just around the corner!

\begin{acknowledgments}
The authors wish to thank  A. Dahlen, R. Easther, B. Freivogel, J.T. Giblin, M. Kleban, A. Lewis, E. Lim, J. Niemeyer, H. Peiris, K. Sigurdson, and P. Steinhardt for helpful conversations. AA was supported by NSF grant PHY-0757912, and a ``Foundational Questions in Physics and CosmologyÕÕ grant from the Templeton Foundation. MJ's research is funded by the Gordon and Betty Moore Foundation, and thanks the Aspen Center for Physics, UC Santa Cruz, the Perimeter Institute, and the Abdus Salam ICTP for their hospitality while portions of this work were completed. 
\end{acknowledgments}

\begin{appendix}
\section{Details of the bubble cosmology}\label{sec:bubblecosmoapp}

In this appendix, we describe a number of the results presented in Sec.~\ref{sec-onebubreview} in greater detail. One key cosmological quantity we require is $R_0$, the physical radius of the observation bubble wall (approximated as a light cone) where it intersects the past lightcone of an observer at $(\xi_o,\tau_o)$. Given this quantity we can determine which region {\em outside} the bubble is in the causal past of the observer.  The radius $R_0$ can be obtained by matching the open coordinates as $\tau\rightarrow 0$ to a set of coordinates outside the observation bubble (see AJ) to obtain:
\begin{equation}\label{eq:defR0}
R_0 =  \lim_{\tau \rightarrow 0} a(\tau) \sinh \left( \int_{\tau}^{\tau_{\rm o}}d\tau/a(\tau) \right).
\end{equation}

The argument of the $\sinh$ is just the elapsed conformal time inside of the bubble, which is determined by the assumed cosmology.  To calculate it, we will first assume that the fields remain potential dominated near $\tau = 0$ (so that inflation occurs at the earliest possible $\tau$, and Eq.~\ref{eq:ainflation} applies) and neglect a period of late--time vacuum energy domination. Then, we will refine this simplified answer. 

It is convenient to split the integral in Eq.~\ref{eq:defR0} into two pieces describing the evolution during and after inflation:
\begin{equation}
\int_{\tau}^{\tau_{\rm o}}d\tau/a(\tau) =  \int_{\tau}^{\tau_{\rm reh}}d\tau/a (\tau) +  \int_{\tau_{\rm reh}}^{\tau_{\rm o}}d\tau/a (\tau),
\end{equation}
where $\tau_{\rm reh}$ is the proper time inside of the bubble at which inflation ends, and the universe reheats. Expanding the $\sinh$ yields:
\begin{eqnarray}
\label{eq-Ro}
R_0 &=&   \lim_{\tau \rightarrow 0} a (\tau) \\
 && \left[ \sinh \left( \int_{\tau}^{\tau_{\rm reh}}d\tau/a (\tau) \right) \cosh \left( \int_{\tau_{\rm reh}}^{\tau_{\rm o}}d\tau/a (\tau) \right)  \right. \nonumber \\
 && \left. +  \cosh \left( \int_{\tau}^{\tau_{\rm reh}}d\tau/a (\tau) \right) \sinh \left( \int_{\tau_I}^{\tau_{\rm o}}d\tau/a (\tau) \right) \right]   \nonumber.
\end{eqnarray}
Now, using the scale factor Eq.~\ref{eq:ainflation} in the integral between $\tau$ and $\tau_I$ and taking the limit as $\tau \rightarrow 0$ gives:
\begin{eqnarray}
&& \lim_{\tau \rightarrow 0}  a_I  \sinh \left( \int_{\tau}^{\tau_{\rm reh}} d\tau/a_I (\tau) \right) = H_{I}^{-1} \tanh(H_I \tau_{\rm reh} / 2) \nonumber \\
&& \lim_{\tau \rightarrow 0}  a_I  \cosh \left( \int_{\tau}^{\tau_{\rm reh}}d\tau/a_I (\tau) \right) = H_{I}^{-1} \tanh(H_I \tau_{\rm reh} / 2). \nonumber
\end{eqnarray}

The total conformal time elapsed between $\tau_{\rm reh}$ and $\tau_o$ can be found directly from the scale factor Eq.~\ref{eq:aradmat} by solving for the conformal time at which $a = 1$, and is given by:
\begin{equation}
\label{eq-conftime}
 \Delta \Xi = \int_{\tau_{\rm reh}}^{\tau_{\rm o}}d\tau/a (\tau) = \log \left[ \frac{1+ \Omega_{r}^{1/2} + \Omega_c^{1/2}}{1+ \Omega_{r}^{1/2} - \Omega_c^{1/2}} \right].
\end{equation}
For $\Omega_r \ll 1$ and $\Omega_c \ll 1$, this is approximately:
\begin{equation}\label{eq:approxxiI}
\int_{\tau_{\rm reh}}^{\tau_{\rm o}}d\tau/a (\tau) \simeq 2 \Omega_c^{1/2}.
\end{equation}
(This is also the approximate distance in $\xi$ out to which a present day observer can see on the reheating surface, providing the approximation given in Eq.~\ref{eq:xireh}.)

Substituting Eq.~\ref{eq-conftime}into Eq.~\ref{eq-Ro} yields
\begin{equation}
R_0 = H_I^{-1} \tanh \left( \frac{N_e}{2} \right) \left[ \frac{1+ \Omega_{r}^{1/2} + \Omega_c^{1/2}}{1+ \Omega_{r}^{1/2} - \Omega_c^{1/2}} \right],
\end{equation}
where $N_e = H_I \tau_{\rm reh}$ is the number of e-folds of inflation inside the bubble. 

While $N_e$ depends on the inflaton potential, we can constrain it observationally by requiring it to produce a given minimal curvature $\Omega_c$. From the Friedmann Equation,
\begin{equation}\label{eq:friedmann}
(1-\Omega_c)^{-1}=1+{3\over 8\pi G\rho a^2},
\end{equation}
and the inflationary scale factor of Eq.~\ref{eq:ainflation},  the curvature density at inflation's end is:
\begin{equation}\label{eq:curvaturereheating}
\Omega_{c, {\rm reh}} = \frac{1}{\cosh^2 N_e}.
\end{equation}
Reheating occurs at very small $\Xi$, where radiation dominates the evolution of the scale factor Eq.~\ref{eq:aradmat}. To match the pre- and post-inflationary metrics, we require that the energy density in curvature remains constant during reheating. Using the Friedmann equation Eq.~\ref{eq:friedmann}, Eq.~\ref{eq:curvaturereheating}, and the metric Eq.~\ref{eq:aradmat} at small $\Xi$, we can then solve for the conformal time at reheating as a function of $N_e$: 
\begin{equation}
\Xi_{\rm reh} = \frac{1}{\cosh N_e}
\end{equation}
During radiation domination, the conformal time can be related to the temperature by (see e.g.~\cite{Bailin:2004zd}) 
\begin{equation}
\Xi_{\rm reh} \simeq \sqrt{45 \frac{\Omega_c}{\Omega_r^{1/2} g^{1/2} }} \left( \frac{10^{-11} {\rm GeV}}{T_{\rm reh}} \right)
\end{equation}
where $g$ is the total number of relativistic degrees of freedom during reheating (expected to be of order $100$ in the MSSM or SUSY GUT models). Equating these relations, the $N_e$ necessary to produce an observed curvature component can be found. The WMAP 5 year data~\cite{Komatsu:2008hk} indicates that the bound on curvature is $\Omega_c < .0081$ within $95 \%$ confidence level and $\Omega_r \simeq 8 \times 10^{-5}$. This requires that 
\begin{equation}
N_e > {\rm arccosh} \left[ .5 \frac{T_{\rm reh}}{10^{-11} {\rm GeV}} \right]
\end{equation}
which for $1 {\rm TeV} \leq T_{\rm reh} \leq 10^{16} {\rm GeV}$ ranges between $30 < N_e < 62$. Given the reheat temperature, changing the total number of efolds slightly  will change the observed curvature component drastically. Over this entire range of scenarios, we have that $R_0 \simeq H_I^{-1}$.

One can now refine these estimates by including a late-time cosmological constant dominated epoch. Integrating the de Sitter metric to obtain an elapsed conformal time, this would correct the total conformal time between reheating and the present by approximately
\begin{equation}
\Delta \Xi = \Omega_c^{1/2} \left[ 1- e^{-H_{\Lambda} (\tau_o - \tau_{\Lambda} )}  \right],
\end{equation}
where we have used the fact that $(a H_{\Lambda})^{-2} \approx \Omega_c$ at the onset of cosmological constant domination. Since $\Delta \Xi \le \Omega_c^{1/2}$ even as $\tau_o \rightarrow \infty$, including this epoch changes out estimate of $R_0$ little. (Note also that the curvature component will monotonically decrease during this epoch, and so $\Omega_c$ today will be the maximum value of the curvature after the original epoch of inflation.)

More interesting for the causal structure of the bubble spacetimes is to include an epoch of curvature domination before inflation begins. This will occur whenever the field energy is not potential dominated. If the initial epoch of curvature domination lasts for a time $\tau_c$ (so that $a = \tau$ from $0 < \tau < \tau_c$) , and we take the scale factor during inflation to be $a = \tau_c + H_I^{-1} \sinh \left[ H_I (\tau - \tau_c)\right]$, then Eq.~\ref{eq-Ro} yields
\begin{eqnarray}
R_0 &=& \tau_{c} \left[ \frac{1+ \Omega_{r}^{1/2} + \Omega_c^{1/2}}{1+ \Omega_{r}^{1/2} - \Omega_c^{1/2}} \right]  \\ &&\times \exp \left[ \int_{\tau_c}^{\tau_I} \frac{d\tau}{H_I^{-1} \sinh \left[ H_I (\tau - \tau_c)\right] + \tau_c} \right] \nonumber
\end{eqnarray}
For $\tau_c > H_I^{-1}$, $R_0$ can be somewhat larger than Eq.~\ref{Rosimp}. In this case, the integral in the exponent turns out to be positive and order one; $R_0$ therefore grows like $\tau_c$. 

\section{Observed angular scale}\label{sec:obsang}
In this appendix, we derive Eq.~\ref{eq:costhetadS} for the observed angular scale of a bubble collision on the reheating surface. The hyperbolic foliation of the bubble interior during an inflationary phase is given by 
\begin{widetext}
\begin{equation}
ds^2 = - (1+H_I^2 z^2) dz^2 + (1+H_I^2 z^2) dx^2 + z^2 \left( d\chi^2 + \sinh^2 \chi d\phi^2 \right).
\end{equation}
This is induced by the embedding space coordinates
\begin{eqnarray}\label{eq:embeddinghyperopen}
&& X_{0} = z \cosh \chi = H_{I}^{-1} \sinh (H_I \tau) \cosh \xi \\ \nonumber
&& X_{1} = H_I^{-1} \sqrt{1+H_I^2 z^2} \sin (H_I x) = H_{I}^{-1} \sinh (H_I \tau) \sinh \xi \cos \theta \\ \nonumber
&& X_{2} = z \sinh \chi \cos \phi = H_{I}^{-1} \sinh (H_I \tau) \sinh \xi \sin \theta \cos \phi \\ \nonumber
&& X_{3} = z \sinh \chi \sin \phi = H_{I}^{-1} \sinh (H_I \tau) \sinh \xi \sin \theta \sin \phi \\ \nonumber
&& X_{4} = H_I^{-1} \sqrt{1+H_I^2 z^2} \cos (H_I x) = H_I^{-1} \cosh (H_I \tau),
\end{eqnarray}
\end{widetext}
where we have also indicated the embedding coordinates for the open foliation. We will want to consider these coordinates in different frames, and a useful relation is:
\begin{equation}
\label{eq-boostinv}
\sinh^2 (H_{I} \tau) = H_I^2 z'^2 - \left( 1 + H_I^2 z'^2 \right) \sin^2 (H_I x').
\end{equation}
Since $\tau$ does not change under a boost, the quantity on the right hand side is boost-invariant. 

In the collision frame, where the bubbles nucleate simultaneously, the motion of a null disturbance in the $\pm x'$ direction obeys
\begin{equation}
\label{eq-nulltraj}
H_I x' = \pm \arctan (H_I z') + C,
\end{equation}
where $C$ is a constant. If the ingoing ($-x'$) null shell originates at $\{z' = z_c , x' = \arctan z_c \}$, where $z_c$ is the location of the collision (we suppress the prime, but this quantity is always understood to be evaluated in the collision frame), the constant is given by $C= 2 \arctan (H_I z_c)$.

Substituting this into the boost-invariant relation eliminates $x'$:
\begin{equation}
\sinh^2 (H_I \tau) = \frac{4 H_I^2 z_c \left( z' - z_c + H_I^2 z_c z'^2 - H_I^2 z_c^2 z' \right)}{\left( 1+ H_I^2 z_c^2\right)^2}, 
\end{equation}
and solving for $z'$ yields:
\begin{equation}\label{eq:relzprime}
H_I z' = \frac{-\left( 1 - H_I^2 z_c^2 \right) + \left( 1 + H_I^2 z_c^2 \right) \cosh (H_I \tau)}{2 H_I z_c}.
\end{equation}

Given the trajectory $z'(\tau)$ for the wall in the collision frame, the idea is to boost this into the observation frame, where the observer is at the origin. Then, the intersection of the $\{x(\tau),z(\tau)\}$ 2-hyperbola with the surface of $\tau=const.$ defines how the observable part of the reheating surface is split into affected and unaffected regions. We can then find the intersection of the past light cone of an observer with this surface to obtain the angular scale of the collision. 

To implement this, we go to the observation frame via the the boost:
\begin{equation}
X_{1}' = \gamma \left( X_{1} - v X_{0} \right).
\end{equation}
In $X_{1}'$, using the embedding, we can solve for $\sin (H_I x')$ and substitute this into the boost-invariant relation, yielding
\begin{equation}
H_I^2 \gamma^2 \left( X_{1} - v X_{0} \right)^2 = H_I^2 z'^2 - \sinh^2 (H_I \tau).
\end{equation}
Next we substitute in our relation for $z'$ Eq.~\ref{eq:relzprime} and then use the embedding Eq.~\ref{eq:embeddinghyperopen} to substitute for the observation frame open slicing coordinates. We assume that the observer can see out to some distance $\xi_{\rm reh}$. Finally, solving for $\cos \theta$, and identifying $\theta$ as half of the observed angular scale ($\theta = \psi / 2$), we obtain Eq.~\ref{eq:costhetadS}.

\section{Dynamics of the radiation shell from a bubble collision}
\label{sec-nullshell}

Here we calculate the energy density on null shells of radiation emanating from a bubble collision using the thin-wall matching formalism of Sec.~\ref{sec:thinwallcollisions}. Working in the collision frame during inflation the path of the null radiation shell in the hyperbolic foliation was determined in Sec.~\ref{sec:obsang} by Eq.~\ref{eq-nulltraj}. Then, using Eq.~\ref{eq:relzprime} and setting $H_I \tau = N_e$, we obtain the position in $z$ of the radiation shell after $N$ efolds of inflation. Using Eq.~\ref{energydensity}, the energy density in the radiation shell for $\cosh N_e \gg 1$ is then given by
\begin{equation}\label{eq:sigmaintermediate}
\frac{\sigma_{N_e}}{H_I^3} = M \frac{H_I z_c^2 }{\pi (1+H_I^2 z_c^2)^2 \cosh^2 N_e}.
\end{equation}

This expression is completely general, but in the case of mild collisions, or collisions between identical bubbles, we can substitute the expression Eq.~\ref{Mo} for $M$ into Eq.~\ref{eq:sigmaintermediate} and determine the energy density in the radiation shells at the end of inflation:
\begin{equation}
\label{eq:sigmaapprox}
\frac{\sigma_{{N_e}}}{H_I^3} = \frac{\left( 1 - \frac{H_I^2}{H_F^2}\right)m_p^2 H_F^2 H_I z_c^5}{ 2 \pi (1+H_F^2 z_c^2)  (1+H_I^2 z_c^2) \cosh^2 N_e }.
\end{equation}
Various limiting cases are explored in Sec.~\ref{sec:observingdebris}.

Another effect of interest is the gravitational redshift of photons as they cross the null shell of radiation. The magnitude of this effect depends on the time at which photons cross the shell of radiation, and hence on the cosmology. Solving for the trajectory of the null shell from reheating to last scattering is in general difficult, and so here we find the red-shift experienced during inflation and (following CKL) in a simplified cosmological model where an inflationary epoch is matched onto an empty open universe at reheating.

During inflation, modes crossing the radiation shell at some position $z$ will experience a red-shift given by
\begin{equation}
1-\frac{a_o}{a_{HdS}} = \frac{2M}{z (1+H_I^2 z^2)}, 
\end{equation}
where $a_{o}$ is the metric coefficient defined in Eq.~\ref{eq-aoc} for the affected region of the observation bubble (Hyperbolic Schwarzschild de Sitter), and $a_{HdS}$ is the metric coefficient in the unaffected portion of the observation bubble (the hyperbolic foliation of de Sitter space). Substituting using Eq.~\ref{energydensity}, this can be written as
\begin{equation}
1-\frac{a_o}{a_{HdS}} = \frac{8 \pi}{H_I z} \frac{\sigma_{N_e}}{H_I^3} \left(\frac{H_I}{m_p}\right)^2,
\end{equation}
where this expression should be evaluated at the value of $z$ where the photons cross the null shell on their way to an observer. For $z_c \sim H_F^{-1}$, we can use the results above for a mild collision or a collision between identical de Sitter bubbles to obtain
\begin{equation}
1-\frac{a_o}{a_{HdS}} = \frac{4}{\cosh^3 N_e} \frac{H_I^2}{H_F^2} ,
\end{equation}
which is always small. For $H_F z_c \gg 1$, we obtain
\begin{equation}
1-\frac{a_o}{a_{HdS}} = \frac{8}{\cosh^3 N_e}
\end{equation}
which is larger than the former limiting case, but still very small. Therefore, the red-shift incurred upon crossing the shell of radiation during inflation is exponentially suppressed after a small number of efolds. This will be negligible at reheating, although
such an effect may manifest itself in the red-shifting of perturbations produced early on during inflation, which are now at the largest observable scales.  

In CKL, a toy cosmology was introduced where an inflationary epoch was matched along the reheating surface at $\tau_{\rm reh}$ onto an empty open universe. The energy density in the null shell relative to the background jumps drastically in this case, leading to a much larger red-shift. For photons crossing the shell after reheating, the incurred red-shift is given by
\begin{equation}
1 - a_{HS} = \frac{2M}{z}
\end{equation}
where $a_{HS}$ is the metric coefficient in Hyperbolic Schwarzschild. The red-shift is largest at the time of reheating, and so an upper bound on the effect in this case can be found by evaluating this at the value of $z$ where the null shell intersects the reheating surface
\begin{eqnarray}
\frac{2M}{z} &\simeq& \frac{H_I^2}{H_F^2} \frac{1}{\cosh N_e}, \ \ \ z_c \simeq H_F^{-1},\nonumber \\
\frac{2M}{z} &\simeq& \frac{H_I^2 z_c^2}{\cosh N_e}, \ \ \ z_c \gg H_I^{-1}
\end{eqnarray}
For this to be appreciable, $z_c$ must be exponentially larger than $H_I$. With a more realistic treatment of the cosmology, we expect the true magnitude of the red-shifting to lie somewhere between this result and the result found above for the red-shifting experienced during inflation, since the presence of any background energy density decreases the size of the effect (as evidenced by the difference between these two results). Because of the exponential suppression with $N_e$, and the improbability of experiencing a large $z_c$, red-shifting of CMB photons across the null shell should therefore always be a negligible effect.

\section{Calculations of probabilities}

\subsection{Number of nucleations to the past}
\label{sec-nucpastcalc}
In Sec.~\ref{sec-expnum}, we discussed the number of collisions expected under various assumptions. Here, we briefly describe how these numbers can be derived. Using the coordinates in Eq.~\ref{eq:outsidehypercoords}, the four-volume element is given by
\begin{equation}\label{eq:dV4}
dV_4 = H_F^{-4} \frac{\cosh^2 \Psi}{\cosh^4 \Upsilon} d\Psi d \Upsilon d\Omega_2^2.
\end{equation}
Integrating this volume element over the various shaded regions in Figs.~\ref{fig-volumes} and~\ref{fig-volumes2}, we obtain Equations~\ref{eq:numbernoboost},~\ref{eq-simpleprob}, and~\ref{eq:numberintersect}. 

We begin by calculating the total number of collisions in the past light cone of an observer, Eq.~\ref{eq:numbernoboost} and~\ref{eq-simpleprob}. The relevant four-volume is between the bubble wall at $\{ \Psi = \infty, \Upsilon = -\infty \}$, the observer's past light cone as per Eq.~\ref{eq:PLCinFV}, the initial value surface, and the boundary of the coordinates $\{ \Psi = -\infty, \Upsilon = -\infty \}$. We will examine two limits, first treating the case where $\xi_o =0$, and then considering the limit where $\xi_o \rightarrow \infty$. 

For $\xi_o=0$, spherical symmetry allows simple integrating over the solid angle, and the initial value surface is $\Psi=\Upsilon$; this in turn gives a maximal value $\Upsilon = {1\over 2}\log[H_F R_0]$ when combined with the lightcone. Integrating over $\Psi$, and then $\Upsilon$ gives:
\begin{widetext}
\begin{eqnarray}\label{eq:case1v4}
dN &=& 4 \pi \lambda H_F^{-4} \frac{d \Upsilon}{\cosh^4 \Upsilon} \left( \int_{\Upsilon}^{-\Upsilon +  \log \left[ H_F R_0 \right]} d\Psi \cosh^2 \Psi \right) \nonumber \\
  &=& \pi \lambda H_F^{-4}  \int_{-\infty}^{\frac{1}{2} \log \left[ H_F R_0 \right]} \frac{d \Upsilon}{\cosh^4 \Upsilon} \left( -4 \Upsilon +2  \log \left[ H_F R_0 \right] -\sinh (2 \Upsilon) - \sinh \left( 2 \left[ \Upsilon -  \log \left[ H_F R_0 \right] \right] \right) \right), \nonumber \\
N &=& \frac{4 \pi \lambda H_F^{-4}}{3} \left( H_F^2 R^2_0 + 2 \log \left[ 1+ H_F R_0\right] \right). 
\end{eqnarray}
\end{widetext}
We can use Eq.~\ref{Rosimp} to estimate $R_0 \simeq H_I^{-1} \tanh \left[ \frac{N}{2}\right]$, yielding
\begin{equation}
\label{eq-nxi0}
N \simeq \frac{4 \pi \lambda}{3H_F^{4}} \left[ \frac{H_F^2}{H_I^2} \tanh^2 \left[ \frac{N}{2}\right] + 2 \log \left[ 1+ \frac{H_F}{H_I} \tanh \frac{N}{2}  \right] \right].
\end{equation}
The enhancement by $(H_F/H_I)^2$ of the number of collisions arises because the matching radius $R_0$ becomes large when $H_I \ll H_F$. 

If $\xi_o \neq 0$, the core idea is the same but the integration is more complicated because after the boost to the observation frame, the initial-value surface is $\theta-$dependent. Generalizing the result of GGV, in the limit of large $\xi_o$, the expected nucleation number approaches
\begin{equation}\label{eq:case1v4largexi}
N \simeq \frac{4 \pi \lambda}{3 H_F^{4}} \left( \frac{H_F^2}{H_I^2} \right) \xi_o.
\end{equation}
While the four volume accessible to an unboosted observer (Eq.~\ref{eq:case1v4}) is generally bounded if there is a cosmological constant in the true vacuum, the four volume available to an observer at large $\xi_o$ is unbounded regardless of the cosmology inside of the observation bubble. This enhancement comes from the region near past infinity that becomes included to the future of the initial value surface in the vicinity of $\theta = 0$. 

It is also possible to build in some assumptions about the observability of collisions into the counting. For example, we might consider only those collisions where the null shell of debris intersects an observable portion of the reheating surface or surface of last scattering (left panel of Fig.~\ref{fig-volumes2}.) The  surface of interest is located at $\tau=\tau_{\rm view}$. The observer can see out to some radius $\xi_{\rm view} $, if this is the reheating surface or surface of last scattering (as we assume in the main text), the estimated in Eq.~\ref{eq:xireh} to be on the order of $\xi_{\rm reh} \sim 0.18$. Following the null geodesics from the boundary of the observable region of the reheating surface back towards the bubble wall, 
\begin{equation}
\xi_{\rm reh} - \xi = \pm \log \left[ \frac{\tanh \left( \frac{N_e}{2}\right)}{\tanh \left( \frac{H_I \tau}{2} \right)} \right].
\end{equation}
The radii at the intersections with the bubble wall are 
\begin{equation}
R_0^{\pm} = H_I^{-1} \tanh \left( \frac{N_e}{2}\right) e^{\pm \xi_{\rm view}}.
\end{equation}
Following these null rays into the false vacuum using Eq.~\ref{eq:PLCinFV} defines the `upper' and `lower' boundaries of the region available for the nucleation of colliding bubbles as depicted in the left panel of Fig.~\ref{fig-volumes2}. 

As can be seen in Fig.~\ref{fig-volumes2}, the addition of the `lower' boundary excises the region near past infinity that becomes important in the limit of large $\xi_o$ and that lead to the enhancement in Eq.~\ref{eq:case1v4largexi}. Therefore even at $\xi_o \neq 0$, for the present calculation we can get a good approximate result by calculating the four-volume available to a $\xi_o=0$ observer. First, we integrate the volume element over $\Psi$ between the two past light cones: 
\begin{equation}
dN = 4 \pi \lambda H_F^{-4} \frac{1}{\cosh^4 \Upsilon} \int_{-\Upsilon + \log \left[ H_F R_0^-\right]}^{-\Upsilon + \log \left[ H_F R_0^+ \right]} d\Psi \cosh^2 \Psi.
\end{equation}
Taking the rather ugly result, we can integrate $\Upsilon$ between $-\infty < \Upsilon < \infty$. This will include some volume to the past of the initial value surface, but in the limit of small $\xi_{\rm ls}$ and $\tau_o \gg H_F^{-1}$, will reproduce the total four-volume to within a factor of $2$. The result is:
\begin{widetext}
\begin{equation}
N = \frac{8 \pi}{3} \lambda H_{F}^{-4} \left[ \log \left[ \frac{R_0^+}{R_0^-} \right] + \sinh \left( 2 \log \left[ H_F R_0^+ \right] \right) - \sinh \left( 2 \log \left[ H_F R_0^- \right] \right) \right].
\end{equation}
Substituting with $R_0^{\pm}$ and simplifying yields:
\begin{equation}\label{eq:Ncase2}
N = \frac{8 \pi}{3} \lambda H_F^{-4} \left[ \frac{H_F^2}{H_I^2} \tanh^2 \left[ \frac{N_e}{2}\right] \sinh (2 \xi_{\rm view}) + 2 \xi_{\rm view} + \frac{H_I^2}{H_F^2} \frac{\sinh (2 \xi_{\rm view})}{\tanh^2 \left[ \frac{N_e}{2}\right]}  \right].
\end{equation}
\end{widetext}
For large $\xi_{\rm view}$, this will be an overestimate for observers at small $\xi_o$, since much of the four volume counted above will be to the past of the initial value surface. To leading order in the small-$\xi_{\rm view}$ limit, and using Eq.~\ref{eq:approxxiI} in the case where $\xi_{\rm view} = \xi_{\rm ls}$, this is approximately
\begin{equation}
N = \frac{32 \pi  \lambda}{3  H_F^{4}} \left( \frac{H_F^2}{H_I^2} \right) \sqrt{\Omega_c}
\end{equation} 
The number of collisions that intersect the observable portion of $\tau_{\rm ls}$ is related to the total number of collisions for an unboosted observer roughly by a factor of $\sqrt{\Omega_c}$, which is roughly bounded by $~ 0.2$. Thus, even for a $\xi_o = 0$ observer, the majority of collisions in the observer's past light cone do not intersect the $\tau = \tau_{\rm ls}$ surface.

As another example, FKNS considered any situation where the post-collision domain wall crosses the worldline of an observer to be incompatible with satisfactory cosmological evolution, and so did not include any such bubbles in the count. This study assumed that the post collision domain wall is null and travels into the observation bubble for a time $\tau = H_I^{-1}$, after which it travels away. (The details of how long it takes the domain wall to accelerate away are largely irrelevant, and quantitatively change the answer only when the acceleration is very large.) The relevant four-volume is shown in the right cell of Fig.~\ref{fig-volumes2}. Qualitatively, the case is similar to the previous case (depicted in the left panel); in particular, the enhancement of the number of collisions at large-$\xi_o$ is again absent, and the collisions will be distributed more or less isotropically.  Quantitatively, to include only those collisions where the domain wall does not cross $\xi=0$ in the observation frame, we replace $R_0^-$ in the previous calculation by $R_0^- = H_I^{-1}$. Performing the integration, we find
\begin{equation}
N = \frac{4 \pi \lambda}{3 H_F^{4}}  \left[ \frac{H_F^2}{2 H_I^2} \left( e^{2 \xi_{\rm view}} \tanh^2 \left[ \frac{N_e}{2} \right] -1 \right)  \right],
\end{equation}
which in the large-$N_e$ and small-$\xi_{\rm view} = \xi_{\rm ls}$ limit is given by
\begin{equation}
N = \frac{4 \pi \lambda}{3 H_F^{4}}  \left( \frac{H_F^2}{2 H_I^2} \right)  \sqrt{\Omega_c}.
\end{equation}

\subsection{Distribution of angular scales}
\label{sec-angscalecalc}

We begin with the volume element in terms of the global slicing coordinates
\begin{equation}
\label{eq-globalmetric}
ds^2 = \frac{1}{H_F^2 \cos^2 T} \left[ -dT^2 + \sin^2 \eta d\Omega_2^2 \right], 
\end{equation}
which is given by
\begin{equation}\label{eq:dN}
dN = \lambda H_F^{-4} \frac{\sin \eta_n}{\cos^4 T_n} dT_n d\eta_n d (\cos \theta_n) d \phi_n,
\end{equation}
where the `n' subscript denotes `nucleation'.
We can use this to calculate the distribution of collision sizes by changing variables from $T_n$ to $\psi$ using Eq.~\ref{eq:costhetadS}. Some necessary relations are:
\begin{eqnarray}
\label{eq:coordrelations}
z_c &=& H_F^{-1} \frac{\cos T_n - \cos \eta_n}{\sqrt{\sin^2 \eta_n - \sin^2 T_n}}, \\ \nonumber
\gamma &=&  \frac{\sin \eta_n}{\sqrt{\sin^2 \eta_n - \sin^2 T_n}}, \ \ \ \
v = \frac{\sin T_n}{\sin \eta_n},
\end{eqnarray}
The parameters $H_{I} \tau_{\rm reh} = N_e$ and $\xi_{\rm reh}$ are fixed by the cosmology inside of the bubble as per Sec.~\ref{sec-cosmobubble}. Integrating $\eta_n$ at constant $\psi(\eta_n,T_n)$, we obtain the distribution function:
\begin{widetext}
\begin{equation} \label{dndpsi}
\frac{dN}{d\psi d(\cos\theta_{\rm obs}) d\phi_{\rm obs}} = \frac{dN}{d\psi d(\cos \theta_n) d\phi_n} = \lambda H_{F}^{-4} \left[ \int_{\eta_{\rm min}}^{\eta_{\rm max}} d \eta_n \frac{\sin^2 \eta_n}{\cos^4 (T_n(\psi)) } \left| \frac{\partial T_n(\psi)}{\partial \psi} \right| \right].
\end{equation}
\end{widetext}
The upper limit $\eta_{\rm max}$ is determined by finding the intersection of a line of constant $\psi$ with the bubble wall (defined by $T = \eta$). The lower limit $\eta_{\rm min}$ is determined by finding the intersection of a line of constant $\psi$ with the initial value surface, given by Eq.~\ref{eq:boostedivsurface}. This is a function of $\theta_n$, as per Eq.~\ref{eq:boostedivsurface}, unless $\xi_o = 0$ (with maximal value defined by the intersection with $T = \eta - \pi$) but the dependence is weak when $\xi_{\rm reh}$ is small. It is also necessary to determine the Jacobian, which can in general be accomplished analytically, although the form is not terribly illuminating, and we omit the exact expression here.

\end{appendix}

\bibliography{collisionsreview}

\end{document}